\documentclass[preprint2,longabstract]{aastex}

\def\deg{\hbox{$^\circ$}}
\def\sun{\hbox{$\odot$}}
\def\farcm{\hbox{$.\mkern-4mu^\prime$}}
\def\farcs{\hbox{$.\!\!^{\prime\prime}$}}

\slugcomment{Accepted by the Astronomical Journal.}

\shorttitle{The Accretion Tail of Arp 285}
\shortauthors{Smith et al.}

\begin{document}


\title{Stochastic `Beads on a String' in the Accretion Tail of Arp 285}


\author{Beverly J. Smith}
\affil{Department of Physics, Astronomy, and Geology, East Tennessee
State University, Johnson City TN  37614}
\email{smithbj@etsu.edu}

\author{Curtis Struck}
\affil{Department of Physics and Astronomy, Iowa State University, Ames IA  50011}
\email{curt@iastate.edu}

\author{Mark Hancock}
\affil{Department of Physics, Astronomy, and Geology, East Tennessee
State University, Johnson City TN  37614}
\email{hancockm@etsu.edu}

\author{Mark L. Giroux}
\affil{Department of Physics, Astronomy, and Geology, East Tennessee
State University, Johnson City TN  37614}
\email{girouxm@etsu.edu}

\author{Philip N. Appleton}
\affil{NASA Herschel Science Center, 
California Institute of Technology, Pasadena CA  91125}
\email{apple@ipac.caltech.edu}

\author{Vassilis Charmandaris\footnote{IESL/Foundation
for Research and Technology - Hellas, GR-71110, Heraklion, Greece
and 
Chercheur Associ\'e, Observatoire
de Paris, F-75014, Paris, France}}
\affil{Department of Physics, University of Crete, Heraklion Greece 71003}
\email{vassilis@physics.uoc.gr}

\author{William Reach}
\affil{Spitzer Science Center, 
California Institute of Technology, Pasadena CA  91125}
\email{reach@ipac.caltech.edu}

\author{Sabrina Hurlock}
\affil{Department of Physics, Astronomy, and Geology, East Tennessee
State University, Johnson City TN  37614}
\email{zshh7@imail.etsu.edu}

\and

\author{Jeong-Sun Hwang}
\affil{Department of Physics and Astronomy, Iowa State University, Ames IA  50011}
\email{jshwang@iastate.edu}



\begin{abstract}
We present Spitzer infrared, 
GALEX UV, and SDSS and SARA optical images 
of the peculiar interacting
galaxy pair Arp 285 (NGC 2856/4), and compare with a
new numerical model of the interaction.  We estimate the ages of clumps
of star formation in these galaxies
using population synthesis models, carefully considering the uncertainties
on these ages.
This system contains a striking example of `beads on a string': a series of
star formation complexes $\sim$1 kpc apart.
These `beads' are found in a tail-like feature that is perpendicular to
the disk of NGC 2856, which implies that it was formed from material accreted
from the companion NGC 2854.
The extreme blueness of the optical/UV colors and redness of the mid-infrared
colors implies very young stellar ages ($\sim$ 4 $-$ 20 Myrs)
for these star forming regions.
Spectral decomposition of these `beads' shows excess emission 
above the modeled stellar continuum
in the 3.6 $\mu$m
and 4.5 $\mu$m bands,
indicating either contributions from interstellar matter
to these fluxes or
a second older stellar population.
These clumps have 
$-$12.0 $<$ M$_{\rm B}$
$<$ $-$10.6, thus they are less luminous than most dwarf galaxies.
Our model 
suggests that
bridge material falling into the
potential of the companion overshoots
the companion.  The gas then
piles up at apo-galacticon before
falling back onto the companion, and 
star formation occurs in the pile-up.  There was a time delay
of $\sim$500 Myrs 
between the point of closest approach between the two galaxies and the
initiation of star formation in this feature.
A luminous (M$_{\rm B}$ $\sim$ $-$13.6) extended (FWHM $\sim$ 1.3 kpc)
`bright spot' is visible at the northwestern edge
of the NGC 2856 disk, with an intermediate stellar population
(400 $-$ 1500 Myrs).
Our model suggests that this feature is part of a expanding
ripple-like
`arc' created by an off-center ring-galaxy-like collision between the
two disks.

\end{abstract}





\section{Introduction}
Galaxy evolution is strongly driven by interactions and mergers
between galaxies.
Interactions can
produce tidal tails and bridges
\citep{toomre72},
increase
star formation rates \citep{kennicutt87, bushouse88},
and trigger
the formation of 
young super-star clusters \citep{holtzman92, holtzman96}.
Tidal material
can contribute to the intergalactic medium 
\citep{morris94} and to intergalactic starlight \citep{feldmeier02}.
Gas-rich galaxy mergers can produce
ultra-luminous
infrared galaxies 
\citep{soifer87, smith87, sanders88},
while
concentrations of stars and gas in tidal features may become independent
dwarf galaxies 
\citep{barnes92, elmegreen93}.

The key to understanding these processes
is careful comparison of
multi-wavelength observations of nearby galaxies with 
dynamical models.
Since
interactions and mergers are even more common
at high redshift
than in the local Universe 
(e.g., \citealp{abraham01}),
detailed studies of nearby interacting
systems
are important for interpreting high redshift surveys.
Such studies 
can provide information on the timescale of the interaction,
the history of gas compression in different
regions, 
star formation triggering, dissipation in the gas,
multiple bursts of star formation,
and mass transfer between galaxies
\citep{struck03, struck05, smith05b, hancock07}.
Computer simulations
can provide predictions of the distribution of star formation,
which can be compared to observational results to
estimate the effects of compression strength, duration, and other
factors (e.g., \citealp{struck03}).

To study star formation enhancement in pre-merger interacting systems,
we obtained mid-infrared observations
with the Spitzer telescope \citep{werner04}
for three dozen interacting galaxies selected
from the \citet{arp66}
Atlas of Peculiar Galaxies (the `Spirals, Bridges,
and Tails' (SB\&T) sample; \citealp{smith07}).
We have completed detailed multi-wavelength studies
of three of the galaxy pairs in the SB\&T sample, and have constructed
matching hydrodynamical models of their encounters:
Arp 284 
\citep{smith97, struck03, smith05a},
Arp 107 \citep{smith05b},
and Arp 82 \citep{hancock07}.
A similar study of the interacting pair IC 2163/NGC 2207 was presented 
by \citet{struck05} and \citet{elmegreen06}, while 
Arp 24 was studied by
\citet{chen07}.

In the current paper, we describe a multi-wavelength study
of another of the SB\&T systems,
the interacting galaxy pair Arp 285 (NGC 2856/4), and compare
with a new numerical model of the interaction.
The more northern galaxy in this widely separated pair, NGC 2856,
has a peculiar tail-like feature extending out perpendicular to the disk
(Figure 1).  \citet{toomre72} suggested that this feature  
is material from the southern galaxy NGC 2854, 
which has accreted onto NGC 2856 via the bridge.
The presence of a massive HI counterpart to this tail
and the HI velocity field support this hypothesis
\citep{chengalur94, chengalur95}.
The Spitzer 3.6 $\mu$m $-$ 8.0 $\mu$m broadband infrared color of the
NGC 2856 tail
is the reddest of all the tidal features
in the SB\&T sample
\citep{smith07},
implying a very young stellar population. 

In the current study, we 
investigate star formation in Arp 285 
by combining our Spitzer 
mid-infrared images 
with 
ultraviolet
images from the Galaxy Evolution Explorer (GALEX) mission 
\citep{martin05} and optical images from the Sloan Digitized Sky Survey (SDSS) 
\citep{abazajian03}
and the Southeastern
Association for Research in Astronomy (SARA) 
telescope\footnote{http://astro.fit.edu/sara/sara.html}.
We also compare with the 2MASS Atlas near-infrared images of
Arp 285 \citep{cutri06}.
Arp 285 is relatively nearby,
at a distance of 39 Mpc (H$_0$ = 75 km~s$^{-1}$Mpc$^{-1}$).

\section{Observations}

The Spitzer infrared observations and data reductions are described
in detail in \citet{smith07}.  The data used includes broadband
3.6 $\mu$m, 4.5 $\mu$m, 5.8 $\mu$m and 8.0 $\mu$m images from
the Infrared Array Camera (IRAC; \citealp{fazio04}),
with spatial resolutions of 1\farcs5 $-$ 2\farcs0, a pixel size
of 1\farcs2, and a field of view of 7\farcm8 $\times$ 12\farcm6.
A 24 $\mu$m image of Arp 285 was also obtained 
with the Multiband
Imaging Photometry
for Spitzer (MIPS;
\citealp{rieke04}), however, it has pronounced artifacts from the 
point spread function (see image in \citealp{smith07}).
Because of these
artifacts, this image is 
only useful for determining total galaxian fluxes, not fluxes for individual clumps
or tidal features.  Thus it is not used in this analysis.

Arp 285 was observed 
as part of the Sloan Digitalized Sky Survey (SDSS)
in the ugriz optical
filters 
(effective wavelengths 3560\AA, 4680\AA,
6180\AA, 7500\AA, and 8870\AA, respectively).
These images have a pixel size of 0\farcs40 and a field of view of 13\farcm5 $\times$
9\farcm8.
The two galaxies in the pair are in two different SDSS
fields of view.
The FWHM point spread function is $\sim$1\farcs2, based on stars in the field.

Arp 285 was also observed with the SARA 0.9m optical telescope on 2006 Jan 29,
in partly cloudy weather.
An 1152$\times$770
Apogee Alta
CCD with
a pixel size of 0\farcs64~pixel$^{-1}$ was used, giving a field of view of 
12\farcm3 $\times$ 8\farcm2.
A total of three 600 second exposures were made in a broadband R
filter,
along with seven 600 second images in a redshifted
H$\alpha$ filter centered at 664 nm with a FWHM of 7 nm.
For Arp 285, this filter contains
both H$\alpha$ and the [N~II] $\lambda$$\lambda$6548,6583 line.
The SARA data were reduced in the standard way using the Image
Reduction and Analysis Facility 
(IRAF\footnote{IRAF is distributed by the National 
Optical Astronomy Observatories, which are operated 
by the Association of Universities for Research 
in Astronomy, Inc., under cooperative agreement with the National
Science Foundation.})
software.  Continuum subtraction was accomplished using the scaled 
R band image.

Arp 285 was observed 
in a 
near-ultraviolet
(NUV) broadband filter (1750 $-$ 2800 \AA) 
by 
GALEX as part of the GALEX 
Medium Imaging Survey (MIS)
\citep{martin05}.
The MIS image had a total integration time of 813 sec.
Arp 285 was also observed in the 
far-ultraviolet (FUV) (1350 $-$ 1705 \AA)
as part of the 
GALEX All-Sky Survey \citep{martin05}, 
with a shorter exposure
time
of  
112 
seconds.
The GALEX spatial resolution is $\sim$5$''$,
with a pixel size of 1\farcs5.  The field of view is circular,
with 
a 1.2{\deg} diameter.

The total magnitudes 
for NGC 2854 and NGC 2856
in the various
filters 
are given in Table 1.

\section{The Morphology of Arp 285}

\subsection{NGC 2856}

In Figure 2,
we present
a montage of the UV, optical, and infrared images of NGC 2856, the northern galaxy
in the Arp 285 pair.
In the optical images, a dusty spiral
pattern and a central bar-like feature are seen in the disk.
A series of four clumps are visible along the northern tail
in all of the optical images, except for the u image (only 
two clumps detected)
and the z image (only one clump detected).
These clumps are labeled on the g image in the last panel in Figure 2.
From south to north, the 
separations between the clumps are
7\farcs6 (1.4 kpc), 6\farcs9 (1.3 kpc), and 5\farcs1 (1.0 kpc).
Clumps 1 and 2 have bright unresolved or marginally-resolved
cores (FWHM $<$ 1\farcs3 = 250 pc) in the g image; clumps 3 and 4 are 
fainter in g,
with multiple peaks.

The
Spitzer 
5.8 $\mu$m and 8.0 $\mu$m images are affected by `banding', where bright
point sources (such as galactic nuclei) cause horizontal `bands' in the images
(Spitzer Infrared Array Camera Data Manual, Version 2.0, 2005).
As discussed in \citet{smith07}, 
we corrected for these artifacts by interpolating from nearby clean regions.
Unfortunately, the correction was not perfect for NGC 2856, 
leaving a residual
diagonal
`stripe' across the rotated image near the tail (see Figure 2).
In spite of this, however,
clump 3 is detected in all four Spitzer bands, and clump 2 is detected
at 3.6 $\mu$m and 4.5 $\mu$m.
Clump 1 is not detected in any of the Spitzer filters, while clump 4
has only a marginal detection at 8.0 $\mu$m.
Note that clump 2 is brightest in the SDSS data, but clump
3 is brightest at 8 $\mu$m.

The northern tail is also visible in both the FUV and NUV images.
In the longer exposure NUV image, clumps 1 $-$ 3 are bright,
while clump 4 is marginally detected.
In the short exposure FUV image,
with the low resolution spatial resolution of GALEX
the individual clumps are not well-resolved.
The northern tail is not detected in the 2MASS 
near-infrared images.

In the SARA H$\alpha$ map (Figure 3), only clump 2 is detected in
the northern tail.  
This implies that clump 3, the brightest clump
at 8 $\mu$m, is more extincted.  
This is consistent with the 
$\sim$12$''$ resolution
C Array 
HI map of 
\citet{chengalur94}.  In this map,
two HI peaks are clearly
visible in the northern tail.  The brightest HI peak
is approximately coincident
with clump 1, while the second is near clump 3.
Thus clump 2 may be less extincted than these other clumps.
This is consistent with our analysis of the optical
colors (see Section 4.2).

In the u through 4.5 $\mu$m images of NGC 2856,
a `bright spot' is visible in the northwestern edge of
the disk of NGC 2856 (see Figure 2).
This `bright spot' is also visible in the Arp image (Figure 1),
and is marginally detected in the 2MASS H and K$_s$ images.
However, it is not seen as a discrete source
at 5.8 $\mu$m, 8.0 $\mu$m, or in the FUV, NUV, 
H$\alpha$, or 2MASS J images (Figures 2 and 3).
Unlike the clumps in the northern tail, this `bright spot' is smoothly
extended in the SDSS images, 
with a FWHM $\sim$ 7$''$ (1.3 kpc) in the g filter, 
without a compact core or cores.

Within the inner disk of NGC 2856, bright 8 $\mu$m and 
H$\alpha$ sources are visible
at the ends of the bar and 
the nucleus (Figures 2 and 3).  The bar is asymmetric,
with 
the clump near the southern end of the bar
being brighter than the northern source in both 8 $\mu$m and H$\alpha$.
In the higher resolution
SDSS images, the sources at the ends of the bar are resolved into
2 $-$ 4 peaks separated by 2$''$ $-$ 3$''$ (0.4 $-$ 0.6 kpc).  

In Figure 4, a band-merged approximately true-color optical SDSS image of
NGC 2856 is presented.  This shows that the clumps in the northern tail
are bluer than the
main disk of the galaxy.  The dust features and the spiral pattern
are also visible in this picture.
The northeastern spiral arm is bluer than the southwestern
portion of the disk.  This is also apparent in Figure 2.
In the FUV, NUV, and u
images, the northeastern
portion of the disk is brighter than
the southwestern section, but in the longer wavelength images, the disk is more 
symmetric.  This
suggests that the difference at shorter wavelengths is due
to extinction.
This implies that the northeastern side of NGC 2856 is closest to us. This
is consistent with the sense of rotation indicated by the HI velocity
field \citep{chengalur94}, assuming the northwestern spiral arm is trailing.

A connecting bridge between the two
galaxies is visible in the smoothed g and r SDSS images (see Figure 5),
but is not seen in u, i, or z.
This bridge is aligned with the `bright spot' in the disk.
In Figure 5, 
the northern tail is visible out to $\approx$72$''$ (14 kpc) from the disk.
A 
bend and a sudden drop-off in brightness is evident in this tail
just north of the four bright clumps of star formation.
Another possible faint clump is visible in the smoothed g image 
$\sim$7.9$''$ (1.5 kpc) northwest of clump 4, north of the bend.
This bend and the bridge are also visible in the Arp image (Figure 1).

\subsection{NGC 2854}

In Figure 6, a montage of the UV, optical, and infrared images
of the southern galaxy NGC 2854 is shown.
On-going star formation 
is detected
along the spiral arms and at the ends of the bar.
The base of the northern tail/bridge appears double in the u, g, r, 5.8 $\mu$m,
and 8.0 $\mu$m images.
A series of clumps are visible in the spiral arms in both
the optical and the infrared images, and 8 $\mu$m-bright sources
are seen at the ends of the bar.
For some clumps, there are 1$''$ $-$ 2$''$ offsets between the optical
and 8 $\mu$m peaks; for others, including the nuclear source, there is no 
clear optical peak associated with the 8 $\mu$m source.
In the last panel of Figure 6, we identify eight clumps selected based on
the 8 $\mu$m image.

The SARA H$\alpha$ and R maps of NGC 2854 are presented in Figure 7, with
the H$\alpha$ superimposed on the g and 8 $\mu$m images.
All of the 8 $\mu$m clumps except clump 6 were detected in H$\alpha$.
In addition, possible H$\alpha$ emission is seen associated with 
the western portion of the double bridge.

An approximately true-color optical SDSS image of NGC 2854 is displayed in
Figure 8.  The double bridge structure is visible in this image.
The knots along the northern arm are visible.  The southern
end of the bar is bluer than the northern end, and the southern arm/tail
is bluer than the northern arm.
The southwestern end of the bar is particularly bright in the UV, but
less so in the mid-infrared (Figure 6).
NGC 2854 appears more symmetric in the Spitzer images
than 
at shorter wavelengths.
This implies that the color variations seen in the optical/UV
are due to extinction.  The color variations suggest that the southern side of 
NGC 2854 is the near side, consistent with the HI velocity field of
\citet{chengalur94} and trailing
spiral arms.

The smoothed g image of NGC 2854 is presented in Figure 9.
A faint optical tail is detected, extending 1\farcm8 (20 kpc) to the south,
coincident with the long HI tail seen by \citet{chengalur94}.
This tail is also visible in the smoothed FUV and NUV images.

Approximately 3$'$ due north of NGC 2854, at 9$^{\rm h}$ 24$^{\rm m}$
2.9$^{\rm s}$, 49$^{\circ}$ 14$'$ 41$''$ (J2000), a small angular size
galaxy is visible on the Arp image (Figure 1).  This galaxy is
detected in all of the GALEX and SDSS bands, as well
as the Spitzer 3.6 $\mu$m and 4.5 $\mu$m filters.  At the
present time, no redshift is available for this source, so it 
is unknown whether it is associated with Arp 285.
Its proximity to a very bright star (see Figure 1) prevents a reliable
H$\alpha$ detection.
The magnitudes of this galaxy in the various filters are given in 
Table 1.  The UV/optical colors are quite blue, consistent with
a young stellar population.

\section{Clump Analysis}

\subsection{Photometry }

The positions, SDSS, and GALEX 
magnitudes and Spitzer flux densities of the clumps 
in the NGC 2856 tail and the two disks of Arp 285
are given in Table 2 in R.A. order.
The 
photometry was done using the IRAF 
{\it daophot} routine.
For the tail and the `bright spot',
the positions were determined by eye based
on the g image, while the disk positions
are the 8 $\mu$m peaks.
These positions are marked in the last panels of Figures 2 and 6.
The apertures 
we utilized are given in Table 3, along with the aperture corrections used
to correct to total magnitudes.
Since the clumps in the northern tail are separated by only $\sim$5$''$
$-$ 7$''$, we 
used relatively small apertures for the tail clump photometry.
For the disk clumps, we used larger apertures because of
less crowding, and 
because there are sometimes multiple
optical peaks associated with a single Spitzer source.

For background subtraction, the local galaxian background 
was determined using the
mode in an annulus surrounding the source (see Table 3).
To estimate the uncertainty in the colors of the clumps
due to background subtraction,
in addition to the statistical uncertainties determined
from the rms in the background annuli, in calculating the colors
we added in quadrature a
second uncertainty term,
determined from comparing the 
clump colors obtained with the above method with those
obtained with slightly larger annulus.

We also extracted approximate J, H, and K$_s$ photometry
for the clumps from the 2MASS Atlas images.   
These near-infrared fluxes were not used in the population synthesis
modeling (Section 4.2), but were only used for comparison with the
Spitzer data (Section 4.3), thus we did not include the second
sky annulus.

We also obtained the total FUV and NUV flux for a
25\farcs8 $\times$ 9\farcs7 region containing the four knots.
These magnitudes are also given in Table 2, along with 
the SDSS and Spitzer magnitudes for the same region.
The uncertainties given in Table 2 for this region were calculated as in
\citet{smith07}, including both statistical uncertainties
and the uncertainty in the sky level.
The total g flux in this rectangular region is $\sim$3$\times$
the sum of the g fluxes for the four clumps (see Table 2), thus there
is significant diffuse emission in this tail.

\subsection{Ages}

In Figure 10, 
we plot the SDSS 
g $-$ r colors
for both the tail and disk clumps vs.\
their u $-$ g colors. 
In 
Figure 11,
NUV $-$ g is plotted against
g $-$ r.
Similar plots of 
r $-$ i vs.\ g $-$ r,
i $-$ z vs.\
r $-$ i, 
and FUV $-$ NUV vs.\ g $-$ r
are presented in the Appendix.
These Figures also include
the colors for the 
25\farcs8 $\times$
9\farcs7 
rectangular region that includes the four clumps in the northern tail.

To estimate the ages of these clumps, we calculated theoretical colors
for star forming regions
using version 5.1 of the 
Starburst99 
population synthesis code
\citep{leitherer99}. 
This version
includes the Padova
asymptotic giant branch stellar
models \citep{vaz05}.
These models assume an instantaneous burst with a Kroupa
(2002)
initial mass function (IMF)
and an initial mass range of 0.1 $-$ 100 M$_{\sun}$.  
We calculated colors for a range of ages $\tau$ from 1 Myr to 10 Gyrs.
We used a time step size
of 
$\Delta$$\tau$ = 1 Myrs for 0 $<$ $\tau$ $<$ 1 Gyr,
and 
$\Delta$$\tau$ = 100 Myrs for 1.1 $<$ $\tau$ $<$ 10 Gyr.
The \citet{calzetti94} starburst dust reddening law was assumed.
We also generated models using a Salpeter IMF, and found the colors
differed only slightly from those with the Kroupa IMF.  This is
consistent with earlier studies
\citep{macarthur04}.
To the broadband fluxes, we added in the contributions from H$\alpha$
emission, which can be substantial in the r filter.  For a 1 Myr 
star forming region, H$\alpha$ decreases the r magnitude
by 1.1 magnitude; at 5 Myrs, H$\alpha$ contributes $\sim$0.25 magnitudes
(see Figure 10).

To systematically estimate ages and extinctions for the clumps in
Table 2,
we used 
a 
$\chi^2$ minimization calculation (e.g., \citealp{pas03}) to
determine the fit of the observed colors to that of the models:

$$\chi^2 = \sum^{N}_{i=1}\left(\frac{obs_{i} -
model_{i}}{\sigma_{i}}\right)^2$$

\noindent In this equation, N is the number of colors used in the analysis,
obs$_i$ is the observed color,  model$_i$ is the corresponding model
color, and $\sigma_i$ is the uncertainty in the obs$_i$ color.  
A good fit is indicated by 
$\chi$$^2$ $<$ N.
In these calculations, we did not include filters with non-detections.
In a few cases, it was not possible to
find a good fit when the low
S/N FUV or z fluxes were included.  In these cases, they were 
not used in determining the ages.

To estimate the uncertainties in the best-fit parameters, we used
the $\Delta$$\chi^2$ method \citep{press92} to determine
68.3\% confidence levels for the parameters.
The best-fit parameters and their uncertainties 
are given in Table 4, along with 
the colors used in the fits.
We assumed 
solar
metallicity in these models.
In the Appendix, we give results assuming
1/5 solar metallicity.
The derived age ranges for the two metallicities 
are similar, thus the assumed metallicity has little effect on the
derived ages.
We note that, because 
the clump masses are relatively low (see Section 4.4),
stochastic sampling of the IMF can affect
the age determinations (e.g., \citealp{cervino02}).
We do not include this effect in
our calculations. 

On the color-color plots of the clump colors,
we superimpose the 
solar metallicity models.
Using only optical colors, it is often difficult to distinguish between
reddening due to age and reddening due to extinction 
(see Figure 10).
Fortunately, however, some clumps were detected in the UV,
which in some cases can
constrain the ages further (see Table 4).
The clumps in the tail have very blue optical/UV colors,
thus 
they have both low extinctions and young ages, with
E(B $-$ V) $\sim$ 0.1 and ages $\sim$ 4 $-$ 20 Myrs (Table 4).
Clump 1 is slightly redder than the other clumps, 
implying a slightly older
age and/or a higher extinction.

As noted previously,
in the HI map of 
\citet{chengalur94}, 
two peaks are visible in the tail,
near clumps 1 and 3.
The brightest peak has an HI column density 
$\approx$ 2 $\times$ 10$^{21}$ atoms~cm$^{-2}$, while 
the second has N(HI)
$\approx$ 10$^{21}$ atoms~cm$^{-2}$.
Assuming the standard Galactic N(H)-to-extinction ratio
of N(H)/E(B$-$V) = 5.8 $\times$ 10$^{21}$ atoms~cm$^{-2}$~mag$^{-1}$
\citep{bohlin78}
and neglecting possible molecular gas, these imply
E(B $-$ V) $\approx$ 0.3 to clump 1 and 
E(B $-$ V) $\approx$ 0.15 to clump 3. 
These are consistent with the population synthesis results
(Table 4).

In the optical colors (Figure 10), the nuclei of the two
galaxies (cyan open diamond 3 and open black circle 7)
are quite red, meaning high extinctions and/or age.  
Since they were undetected with GALEX,
we are not able to tightly constrain
their ages (see Table 4).
In the case of the `bright spot' in the NGC 2856 disk, 
the optical colors can 
only constrain the age to between 5 $-$ 1500 Myrs.
The lack of a detection in the NUV, however, further constrains
the age to be 
$^{>}_{\sim}$400 Myrs (see Figure 11).  

We also attempted to model the ages of the diffuse
emission in the two galaxian disks and the northern tail
by subtracting the clump light from the total emission.
We modeled the star formation history
of the diffuse emission
in two ways: as an instantaneous burst
and as continuous star formation.  
With the exception of ruling out
extremely young ages (1 $-$ 5 Myrs),
we cannot strongly constrain the age of the diffuse
emission.
For 
the instantaneous burst models,
we get upper limits to the ages of
the diffuse emission in the NGC 2854 and NGC 2856 disks
of
1.5 Gyrs and 400 Myrs, respectively.
This implies that there are some recently formed stars in the disks
outside of the regions
we defined as clumps.  This does not, however, rule out an additional
underlying older stellar population in the disks.


\subsection{Spitzer Colors}

Figure 12 shows the 
Spitzer [3.6 $\mu$m] $-$ [4.5 $\mu$m] 
vs. [4.5 $\mu$m] $-$ [5.8 $\mu$m] colors for the clumps in the
NGC 2856 tail (magenta
diamonds), the NGC 2856 disk (cyan open diamonds), and the NGC 2854 disk (black
open circles).  
A similar plot for 
[4.5] $-$ [5.8] vs.\ [5.8] $-$ [8.0] is provided in the Appendix.
These colors are compared to the colors
of 
the Arp 107 
and Arp 82 clumps (green diamonds,
from \citealp{smith05b} and \citealp{hancock07}), 
Galactic interstellar dust (blue X's; from \citealp{flagey06}), 
M0III stars (open blue square, from Cohen 2005, private communication),
and field stars (magenta open triangle, from \citealp{whitney04}), as well
as the total colors for the NGC 2856 tail region (red filled triangle).

As show in Figure 12, [3.6] $-$ [4.5] $\approx$ 0.0 for both stars
and Galactic dust.  Global values for both interacting and 
spiral galaxy disks are also close to this value \citep{smith07},
as are the clumps in the disks of Arp 285, 107, and 82 (Figure 12).
The [5.8] $-$ [8.0] colors of most of the clumps 
in the three Arp systems
are similar to those
of interstellar matter and redder than 
stars, as expected since these
bands are likely dominated by interstellar dust emission.
The 
[4.5] $-$ [5.8] colors of the Arp clumps 
are 
mainly
between those of stars
and interstellar matter (Figure 12), suggesting contributions from both.
The very red [4.5] $-$ [5.8] color
of clump 3 in the Arp 285 tail compared to the
clumps in Arp 107 and Arp 82 and the
other clumps in Arp 285 (Figure 12) implies 
more contributions from interstellar matter.
This is consistent with the very young age determined from the optical
colors (Table 4).

To disentangle the contributions from 
starlight and dust to the Spitzer bands, the results of our
stellar population synthesis (Section 5) are helpful.
In Figure 13, we plot the full optical-mid-infrared spectral
energy distribution (SED) for tail clump 3.  We superimpose
on this plot our best-fit Starburst99 model (4 Myrs, E(B$-$V) = 0.1),
along with models that span
our 1$\sigma$ uncertainty (68\% confidence)
in the age (3 $-$ 6 Myrs).
In addition, we include a theoretical dust spectrum from
\citet{dl07}.  This dust spectrum shows the broad polycyclic
aromatic hydrocarbon (PAH) mid-infrared emission features,
as well as a `hot dust' continuum.  The plotted dust
spectrum was calculated with the solar neighborhood interstellar
radiation field, scaled up by a factor of U = 100. It uses
a PAH-to-total-dust mass ratio of q$_{PAH}$ = 4.6$\%$.
The dust SED in our wavelength range does not vary much with U;
lowering q$_{PAH}$ weakens the PAH features \citep{dl07}.
The dust model has been scaled to fit the observed 8 $\mu$m flux.
The solid line in this plot is the combined stellar-dust
spectrum.  
The H$\alpha$ contribution to the SDSS r band is clearly visible in
this plot.  A contribution to 
the broadband 3.6 $\mu$m
Spitzer flux from the 3.3 $\mu$m PAH feature is also apparent.

Figure 13 indicates that, in addition to the PAH contribution to the
3.6 $\mu$m band and the modeled stellar component, 
the 3.6 $\mu$m and 4.5 $\mu$m fluxes of tail clump 3
also include another component.
Either there is 
a significant `hot dust' contribution to these bands, as 
indicated by the \citet{dl07} model, or there
is a second underlying older stellar component that is not revealed
by
the optical/UV population synthesis.

For comparison to tail clump 3, 
in Figure 14 we plot the 
SED of clump 2 in the NGC 2856 disk.
Although this also has a very young stellar population (see Table 4),
it has a much more reddened SED than the tail clump because of much
higher extinction.   Also, although PAH emission is clearly
present at 8 $\mu$m, in the 3.6 and 4.5 $\mu$m bands 
most of the emission is stellar, in contrast to clump 3 in the tail.

The `bright spot' in the NGC 2856 disk (cyan clump 1) is undetected
at 5.8 $\mu$m and 8.0 $\mu$m, with a very blue
[4.5] $-$ [5.8] 
upper limit compared to
the other clumps (Figure 12).
This suggests that this region has an older stellar population than the
other disk clumps.
As noted earlier, 
with the available optical data we could not strongly constrain the age
of this clump (Table 4), however, the lack of an NUV detection
points to an older age (see Section 4.2).   
The Spitzer results are consistent with this conclusion.
This shows that Spitzer mid-infrared data may
be useful for breaking the age-extinction degeneracy
in optical colors.
In Figure 15, we plot the SED for the `bright spot', with the
best fit from the optical data shown.  The NUV limit plotted
shows the additional constraint on the age.
The SED plot shows that the 3.6 $\mu$m and 4.5 $\mu$m
emission is dominated by starlight, with very little if any 
dust contributing.

The two Arp 285 nuclei have 
[4.5] $-$ [5.8] and [5.8] $-$ [8.0] colors similar to the other clumps,
implying nuclear starbursts. 
This is in contrast to 
the Arp 107 nuclei (Figure 12), 
which have older stellar populations \citep{smith05b}.
The two nuclei in Arp 82, like those in Arp 285,
have Spitzer colors of star forming regions
\citep{hancock07}.

\subsection{Absolute Magnitudes and Masses }

In Table 5, we compare the absolute optical magnitudes of the NGC 2856
tail clumps with dwarf galaxies, candidate tidal dwarf galaxies (TDGs),
`super star clusters' (SSCs), and the tail clumps in Arp 82.
The Arp 285 tail clumps are lower luminosity than
most nearby irregular galaxies and tidal dwarf galaxies,
and are
near the lower end of the range for 
SSCs.
The faintest Arp 285 tail clump, clump 4, is somewhat 
less luminous than R136, the
bright star cluster in 30 Doradus in the Large Magellanic Cloud
\citep{oconnell94}.
In contrast, the `bright spot' in the NGC 2856 disk is near the
median for dwarf irregular galaxies.

For the Arp 285 clumps,
in Table 6 we give the range of
stellar masses 
inferred
from the Starburst99 models. 
In this table, we also provide 
stellar masses of various other objects for comparison.
The tail clumps are similar in mass to Galactic globular clusters,
but have lower stellar masses than
those inferred for tidal dwarf galaxies and dwarf irregular galaxies.
The mass of the 
NGC 2856 disk `bright spot' is near the median for dwarf irregular
galaxies.

The 3.6 $\mu$m Spitzer band
is sometimes used as a 
tracer of stellar mass (e.g., \citealp{li07}).  However,
our SED plots (Figures 13 $-$ 15)
show that the stellar mass-to-3.6 $\mu$m
luminosity varies significantly from clump to clump, depending
upon the star formation rate and gas-to-star ratio.
This is illustrated in Figure 16, where we plot the 
stellar mass of the clump determined from the population synthesis
model against the 3.6 $\mu$m luminosity.  
We have also included values for
the clumps in Arp 82 \citep{hancock07}.
On this curve, we have superimposed
lines of constant stellar mass-to-light ratios 
of M/L$_{3.6}$ =
1 M$_{\sun}$/L$_{\sun}$ (solid line) and 
M/L$_{3.6}$ =
10 M$_{\sun}$/L$_{\sun}$ (dotted line), where
L$_{\sun}$ is the bolometric luminosity of the Sun.
This plot shows that the NGC 2856 tail clumps
and four clumps in NGC 2854 (clumps 1, 3, 5, and 6,
in the outer parts of the spiral arms)
have lower 
M/L$_{3.6}$ ratios than the other clumps, which are
close to the 
M/L$_{3.6}$ =
10 M$_{\sun}$/L$_{\sun}$ line.
This indicates that contributions from hot dust
and/or the 3.3 $\mu$m PAH feature
to the 3.6 $\mu$m flux
are significant 
in the tail and outer spiral
arm regions.
Thus caution
should be used in utilizing the Spitzer 3.6 $\mu$m
band to estimate stellar masses in star forming
regions.  For example, for clump 3 in the tail,
the stellar mass is $\sim$1/30th that expected based
on the  
M/L$_{3.6}$ =
10 M$_{\sun}$/L$_{\sun}$ relationship.

\section{A Numerical Model of the Encounter}

To interpret these observational results in terms of the 
dynamical and star forming history of Arp 285,
we have constructed a numerical simulation of the Arp 285 interaction 
using the smoothed particle hydrodynamics (SPH) code of \citet{struck97}.
This code was previously used to model Arp 284 \citep{struck03},
IC 2163/NGC 2207 \citep{struck05}, Arp 107 \citep{smith05b},
and Arp 82 \citep{hancock07}.  

\subsection{Constraints on the Model}

Arp 285 is less 
symmetric than the ring galaxies or planar fly-by encounters 
like M51, NGC 2207/IC2163, and Arp 82. 
The collisional morphology of Arp 285
appears somewhat similar to that of Arp 284, an asymmetric ring/tail 
galaxy (NGC 7714) with an edge-on companion (NGC 7715).
The substantial bridge and tail of NGC 2854, like those of NGC 7715,
lead us to believe that it suffered 
a strong prograde encounter. 
There are also similarities between NGC 7714 and NGC 2856.
The optical images show that the `bright spot' in the northwestern
section of the NGC 2856 disk (disk clump 1) 
is part of an arc-like structure (see Figures 1 and 2).
This arc is 
reminiscent of the partial ring in NGC 7714 (see the \citet{arp66}
photograph of Arp 284), which has
been successfully modeled by an off-center collision \citep{struck03}.
It is also reminiscent of 
the `ripples' in Arp 227, 
which were also modeled by a ring galaxy-like collision 
by \citet{wallin88}.

There are some differences between NGC 2856 and NGC 7714, however.
In contrast to NGC 7714,
NGC 2856 lacks strong tidal tails, 
except for the northern tail perpendicular to the disk and 
a short HI extension to the northwest \citep{chengalur94}.
This suggests that NGC 
2856 did not experience the encounter as very prograde. 
It also does not have the fan-like form common to strong 
retrograde encounters.
This suggests that the orbital path of the two galaxies
is at a large angle to the plane of the NGC 2856 disk.

These considerations give us some idea of the type of 
collision that produced the current morphologies. 
In our simulation of this encounter,
we have 
limited ourselves to the goal of reproducing the large-scale 
morphological structures,
but have not 
attempted to simulate internal disk structures 
nor match the system kinematics
in any 
detail.

One key feature we would like the models to help us 
understand are the beads in the tail north of NGC 2856.
We have considered several conceptual ideas for the origin of this material. The HI morphology suggests that this material is an extension of the bridge from NGC 2854, though the optical observations look as though the bridge curves away from that direction before connecting to the bead region. It may be that the ‘bridge’ is in fact a tidal tail, which is merely projected onto NGC 2856, not connected. However, the HI kinematics indicate that this is unlikely. Moreover, the bead material seems strongly affected by the gravitational potential of NGC 2856. 

Thus, it seems likely that the bead material is accreting onto 
the halo of NGC 2856 from the bridge. There 
two possibilities for how this occurs:
i) as infall through the disk of NGC 2856 and out the other side, or 
ii) by swinging around that disk to the other side. 
It is difficult to distinguish between these two scenarios 
observationally.
In option i) we can imagine that clouds pushing through the 
NGC 2856 disk are shocked and compressed.  This may
trigger star cluster formation, accounting for the beads. 
We would naively expect this process to be sequential, so 
that the beads furthest from the disk are oldest. 
In contrast,
in option ii), a group of inflowing clouds pile up in the halo of 
NGC 2856 and collide with material that arrived earlier.
This could 
trigger star formation simultaneously at several locations. 
Thus option i) would predict an age gradient, while for
option ii)
we would expect roughly coeval clumps.
With the available data, we cannot distinguish between
these two possibilities, since the expected
age gradient for option i) is too small to measure.
Assuming a nominal velocity for the tidal material
away from the disk 
of $\sim$300 km~s$^{-1}$ and motion in the plane of the sky,
for scenario i)
we would expect an age difference
of $\sim$12 Myrs between the first and fourth clumps
in the tail, and $\sim$4 Myrs between clumps 2 and 3.   
This is smaller than the uncertainties on the ages of
these clumps (Table 4).

Another way to distinguish between these two scenarios
is with numerical models of the interaction.
For option i),
we were not able to construct a viable simulation
with a small number of trial runs.
The fundamental difficulty is that in order to 
produce the spirals and other tidal structures in 
NGC 2854 the collision must have a substantial prograde 
fly-by component with respect to NGC 2854.
In that case,
however, material accreted onto 
NGC 2856 from NGC 2854 generally has too much 
relative angular momentum to fall directly onto the NGC 2856 disk. 
Because of this, we suspect that such models occupy a small 
volume of the collision parameter space.
We have therefore chosen to focus on models for option ii).
These are discussed in the next two sections.

\subsection{Model Details}

In the SPH
code, hydrodynamical forces are calculated
on a grid with fixed spacing. Gravitational forces are computed between particles in adjacent cells, to capture local gravitational instabilities. 
The model galaxies have disks containing both gas particles and collisionless star particles, as well as rigid dark halo potentials (see 
\citet{struck97} 
for details). 
Gas particles with densities exceeding a constant density threshold
are identified as star-forming particles.
These generally exceed the local Jeans critical mass.
A number of simulations were run; we will only present the 
results of the best model.

The evolution of our numerical model for the Arp 285 system is 
presented in Figure 17, with additional timesteps provided
in the Appendix.
We adopt the convention that the model primary 
corresponds to the southern galaxy NGC 2854 and the companion to NGC 2856. 
The particles in Figure 17 are color-coded according to their galaxy of
origin, with red particles originating from the primary disk
and green from the companion.
A total of 13,590 star 
and 42,900 gas particles were used in the 
primary disk and 5640 star and 5640 gas particles in 
the secondary disk. 
In this model, the length unit = 1.0 kpc, and the time unit is 200 Myr. 
Figure 17 shows four timesteps in the simulation.
The first plot (top left) shows the appearance in the plane
of the sky near the time
of closest approach, 
where the separation between the two galaxies is $\sim$12 kpc.
The second plot (top right) shows the system
370 Myrs
after
closest approach, while the third (bottom left) shows its
appearance 510 Myrs after closest approach.
These two plots match approximately the observed appearance
at the present time.
The last plot shows the appearance 740 Myrs after closest approach,
the predicted appearance in the future.

The radii of the primary star and gas disks are 
6.0 and 10.8 kpc, respectively. The companion star and gas disk radii are both 3.6 kpc. 
The primary disk was set up in the x-y plane. The companion disk is first set up in the x-y plane, then rotated $40^{\circ}$ around a y-axis through its center, and then $–90^{\circ}$ around the z-axis passing through its center. The relative orbit of the companion is in the x-y plane, so from 
the point of view of the companion disk, the primary approaches at a fairly steep angle.  
In the companion disk of the model in Figure 17, the south side is
the near side.

The orbit is counter-clockwise, as is the rotation of the primary, so it sees the encounter as very prograde. The companion disk rotation, in the x-y plane before the tilts are applied, is clockwise. The initial (x,y,z) position of the companion relative to the primary center is  ($-$8.9, $-$20.0, 0.0) 
kpc. Its initial relative velocity 
is (250, 75, 0) km~s$^{-1}$.

The form of the halo potential of the two galaxies is such that the acceleration of a test particle in this halo is 

\begin{equation}
a =
\frac{G{M_h}}{{\epsilon}^2}\
\frac{r/{\epsilon}}{(1 + r^2/{\epsilon}^2)^{n_h}},
\end{equation}

\noindent
where $M_h$ is a halo mass scale, $\epsilon$ is a core radius (set to 2.0 and 4.0 kpc for the primary and companion, respectively), and the index $n_h$ specifies the compactness of the halo. For the primary we 
use $M_{h} = 1.3 \times 10^{10} M_{\odot}$ and $n_h = 1.2$, which gives a slightly declining rotation curve at large radii. For the companion we take $M_{h} = 2.8 \times 10^{10} M_{\odot}$ and $n_h = 1.35$, 
which gives a more rapidly declining rotation curve. The 
model includes the effects of dynamical friction with a Chandrasekhar-like frictional term (see \citealp{struck03}). 
The effects of this term are small except near closest approach.

With these potentials, the halo masses for the primary and 
companion out to a radius of 12 kpc 
(about the separation at closest approach) are 3.7 $\times$ $10^{10}$ 
M$_{\odot}$ and 
3.4 $\times$ $10^{10}$ M$_{\odot}$, respectively,
with a ratio of about 0.92. This 
is in accord with the near equality of the r and i band 
luminosities of the two galaxies (Table 1).

\subsection{Model Results}

The general morphology of the system is quite well reproduced by the model, including the moderate countertail on NGC 2854
and the bridge (see Figure 17). 
A very close encounter is required to produce a bridge as 
massive as observed. 
On the other hand, the moderate-sized tail of the primary galaxy is the result of a prograde perturbation that was not prolonged. These facts, and the relatively large separation between the galaxies, argue that the relative orbit of the companion is quite elliptical, as in the model. 

The model primary disk is more circular in appearance than that of NGC 2854. There are several possible reasons for the difference. The first is simply that the model disk should have a greater tilt relative to the plane of observation (here the x-y plane). 
The primary disk in the model is in the x-y plane of the sky.
However, as noted in Section 3.2, based on extinction arguments and
the HI velocity field,
the real disk is somewhat
inclined to the line of sight, with the south side closest.  
It is also possible that tidal stretching is responsible for the shape of the primary disk. However, in that case we might expect a longer and more massive tidal tail. This is a rather soft argument at present, but it does appear that the 
bar and the spiral arms of the primary disk are disproportionately strong relative to the tail. This suggests that the bar and spiral arms were present in 
the NGC 2854
disk before the encounter. This possibility was not included in the modeling. 

In addition to the bridge, the model companion galaxy has two
tidal tails, one made of material originating from the companion galaxy itself,
and one from material accreted from the primary galaxy along the bridge
(see Figure 17).
The 
tidal plume drawn off the companion 
disk is visible as the green
feature extending northward in the last three panels of Figure 17. 
We equate 
this structure 
with the HI emission extending to the northwest in the
\citet{chengalur94} HI maps, 
though it is not at the same position angle as in the observations,
being oriented about 45$^{\circ}$ too much to the north compared to
the data.
The red feature extending to the northeast in Figure 17 
we associate with the northern HI tail containing
the `beads' of star formation.
As with the other feature, the position angle
of the model tail is somewhat
off from the observed orientation.

In the model, the disk of the companion was
tilted relative to the direction of the encounter, 
so the perturbation had both an 
orthogonal, ring-galaxy-like component, 
and a retrograde component. 
Waves with circular arc-like components 
develop in the disk of the companion.
This behavior might account for the 
northwest arc-like structure in NGC 2856
containing the `bright spot'.

As with NGC 2854,
the observed structure of NGC 2856 also shows a bar and internal arms. 
However, in this galaxy the structure of the bar 
is rather irregular. 
The simulation shows that a large mass of gas loses 
angular momentum 
as a result of the encounter, and forms 
a compressed inner disk or bar. 
Thus, the bar in NGC 2856 may be the result of the collision, and
may not have existed before the encounter.

The evolution of the bridge in this model 
is especially interesting. Because of its elliptical 
trajectory, the companion speeds past its point of closest 
approach as the bridge begins to form. As the bridge 
initially stretches outward from the primary center, 
it lags behind the companion. Later, the companion 
nears its apogalacticon relative to the primary, and slows, 
so the bridge catches up to it. The bridge material has 
significant angular momentum relative to the companion center, 
so the outermost points swing around to the far side of the companion. 
Shortly thereafter the bridge material
begins to pile-up at an outer radius 
northeast of the companion. 
As time goes on, more bridge material 
streams 
into this pile-up region,
and 
compression 
drives star formation. 
In 
Figure 17, star-forming gas particles are marked with cyan astericks. 
In the second panel, one asterisk is visible in the pile-up region. 
Comparison of different timesteps and different models 
shows that the star formation there is quite stochastic. 
Sometimes there are a number of star-forming particles there, 
and occasionally they line up like the observed `beads'. 
Since the model does not accurately represent 
the effects of self-gravity across this 
pile-up region, the real environment 
may trigger more such star formation than in the model.

Material from both the accretion tail and the companion's
plume 
eventually
accrete onto the companion.
Gas in the companion is compressed by 
the tidal perturbation, and experiences prolonged accretion. 
In the model, the density threshold for star formation 
is easily exceeded, and central star formation continues for
some time. 
This is consistent with
the uncertainties on the age of the stellar population 
in disk clump 3 (the nucleus) of NGC 2856 (Table 4).

Our simulation
somewhat resembles models of polar ring formation via
accretion 
from a companion \citep{resh06},
however, the two
model galaxies merge
before this polar ring proceeds
very far in its development (see later timesteps in Appendix).
The star clusters formed 
in the pile-up region will eventually be 
carried with the companion halo into the merger with the primary. 
They are likely to end up orbiting in the inner halo of 
the merger remnant and possibly adding to the globular cluster 
population there. This is in contrast to dwarf galaxies 
formed at the end of tidal tails, which may spend 
long periods in the outer halo.

\section{Discussion}

The clumps in the northern tail of NGC 2856
are a striking example of the `beads on a string' phenomenon,
in which star forming regions are regularly spaced $\sim$1 kpc
apart along spiral arms and in tidal features 
\citep{elmegreen96}.
Such a `beads on a string' morphology may be indicative of the gravitational
collapse of interstellar gas clouds under self-gravity 
\citep{elmegreen96}.
Similar `bead strings' are also seen in 
the interacting systems IC 2163/NGC 2207
\citep{elmegreen06}
and Arp 82 \citep{hancock07}.  
In IC 2163/NGC 2207, 
the `beads' resolve
into associations of star clusters in higher resolution Hubble Space
Telescope (HST) images \citep{elmegreen06}.
In general, HST images of nearby
galaxies show that young star clusters
themselves tend to be clustered into complexes with 
typical sizes of $\sim$1 kpc \citep{zhang01, larsen04, bastian05}.
As noted earlier, clumps 3 and 4 in the NGC 2856 tail have multiple peaks
visible in the SDSS images.
It is possible that the other two tail clumps,
clumps 1 and 2, which
are unresolved in the SDSS
images, 
will also resolve into multiple star clusters at higher
resolution.

The optical-UV colors of the clumps 
in the NGC 2856 tail are very blue,
and 
imply ages of only $\sim$4 $-$ 20
Myrs.
This is much younger than the time since the point of closest
approach between the two galaxies,
showing that there is a time delay between the initiation of star 
formation and the time of closest approach between the two galaxies.
According
to our numerical model of this system, there should be 
an underlying older stellar component in this tail, made of 
stars stripped from the NGC 2854 disk.
Diffuse optical light is clearly present between the clumps,
and the stellar tail extends 41$''$ (7.8 kpc) to the north
beyond clump 4.  However, we are not able to tightly
constrain the age of this diffuse
stellar population.  Thus it is unclear from the available data
whether
a stellar component to the tail existed
before the current star forming episode.

The tail clumps are lower mass than concentrations in other
tails previously classified as tidal dwarf galaxies (see Table 6).
They are more similar in mass to globular clusters than dwarf
irregular galaxies.  Because of their low mass and the lack of
24 $\mu$m detections and calibrated H$\alpha$ measurements,
it is not possible to get accurate star formation rates
for these clumps.  Very roughly, using the 8 $\mu$m
luminosity 
for 
clump 3 in the northern tail, 
and assuming the 
8 $\mu$m $-$ 24 $\mu$m
relationship found for M51 clumps of \citet{calzetti05} and their
correlation between star formation rate 
and 24 $\mu$m luminosity, we find a
star formation rate for this clump of $\sim$10$^{-3}$ M$_{\sun}$~yr$^{-1}$.
This value is very uncertain due to the low mass
and the bootstrapping from the 8 $\mu$m flux.

In our model, gas from the bridge falling into the potential
of the companion overshoots the companion, piling up in an accretion
tail on the far side of the companion.
Star formation occurs in this region.   Our model suggests that 
the `beads on the string'
may be the result of stochastic processes, albeit in a density
enhanced pileup zone.
It is
possible
that local self gravity is pulling clumps together. The spacing between
the star forming regions in the model
is comparable to the scale of local self-gravity in the code.
At most timesteps, the star formation is found in a couple of
isolated
clumps,
without any `beads' appearance.
Thus it appears we are seeing this feature at a favorable time.

The Arp 285 tail is not unique.
Accretion from a companion along a bridge
may have 
produced
the
star forming
`countertails' 
in Arp 105 \citep{duc94, duc97}
and Arp 104 \citep{roche07}.
In addition,
the inner tail on the western side of NGC 7714, which also
has strong star formation \citep{smith97},
may have formed from accretion from the companion
\citep{struck03}.

Our model suggests that the so-called `bright spot' in the northwestern
portion of the NGC 2856 disk, and its associated arc, were likely 
caused by a ring-like perturbation of the disk by an encounter 
which was mainly perpendicular to the plane of the NGC 2856 disk.
The age of the stellar population in this region is estimated
to be between 400 $-$ 1500 Myrs, while the interaction model
indicates that the point of closest approach between the galaxies 
occurred between about 300 $-$ 500 Myrs ago.   This is consistent 
with the idea that the brightness of this `spot' 
may be due to past star formation
triggered by the encounter.

\section{Summary}

We have investigated star formation in the interacting galaxy
pair Arp 285 using Spitzer infrared, GALEX ultraviolet, and ground-based
optical data, and have constructed a numerical model
of the interaction.  The northern galaxy in this
pair contains an unusual tail-like feature extending perpendicular
to the disk.  Our model suggests that this structure was created
by gas from the companion falling into the gravitational potential
of the disk and overshooting the disk.

A series of regularly-spaced knots of recent star formation are seen
in this tail.  Stellar population synthesis suggests that
these knots have ages of $\sim$4 $-$ 20 Myrs and
masses 
in the range of globular clusters.
The Spitzer 3.6 and 4.5 $\mu$m fluxes from these tail clumps
are higher than expected from the population synthesis,
indicating that either a second older 
stellar population
is present, or there are significant contributions
to these bands from
hot dust.

The `bright spot' in the NGC 2856 disk has an intermediate-age
stellar population (400 $-$ 1500 Myrs).  This feature
and its associated arc
may have been caused by a ring-like disturbance from an encounter
almost perpendicular to the plane of the disk.
Its brightness might be due to past star formation triggered
by the interaction.

\acknowledgements







\acknowledgments

We thank the Spitzer, GALEX, and SDSS teams for making this research possible.
This research was supported by NASA Spitzer grant 1263924, NSF
grant AST-0097616, NASA LTSA grant NAG5-13079
 and NASA GALEX grant GALEXGI04-0000-0026.
V. C. acknowledges partial support from the EU ToK grant 39965.
We thank Jayaram Chengalur for providing us with an electronic copy
of the HI data.   We also acknowledge Amanda Moffett and Chris Carver
for help with system administration.
This research has made use of the NASA/IPAC Extragalactic Database (NED) which is operated by the Jet Propulsion Laboratory, California Institute of Technology, under contract with the National Aeronautics and Space Administration.

\clearpage



 \begin{figure*}
\includegraphics[width=6.3in]{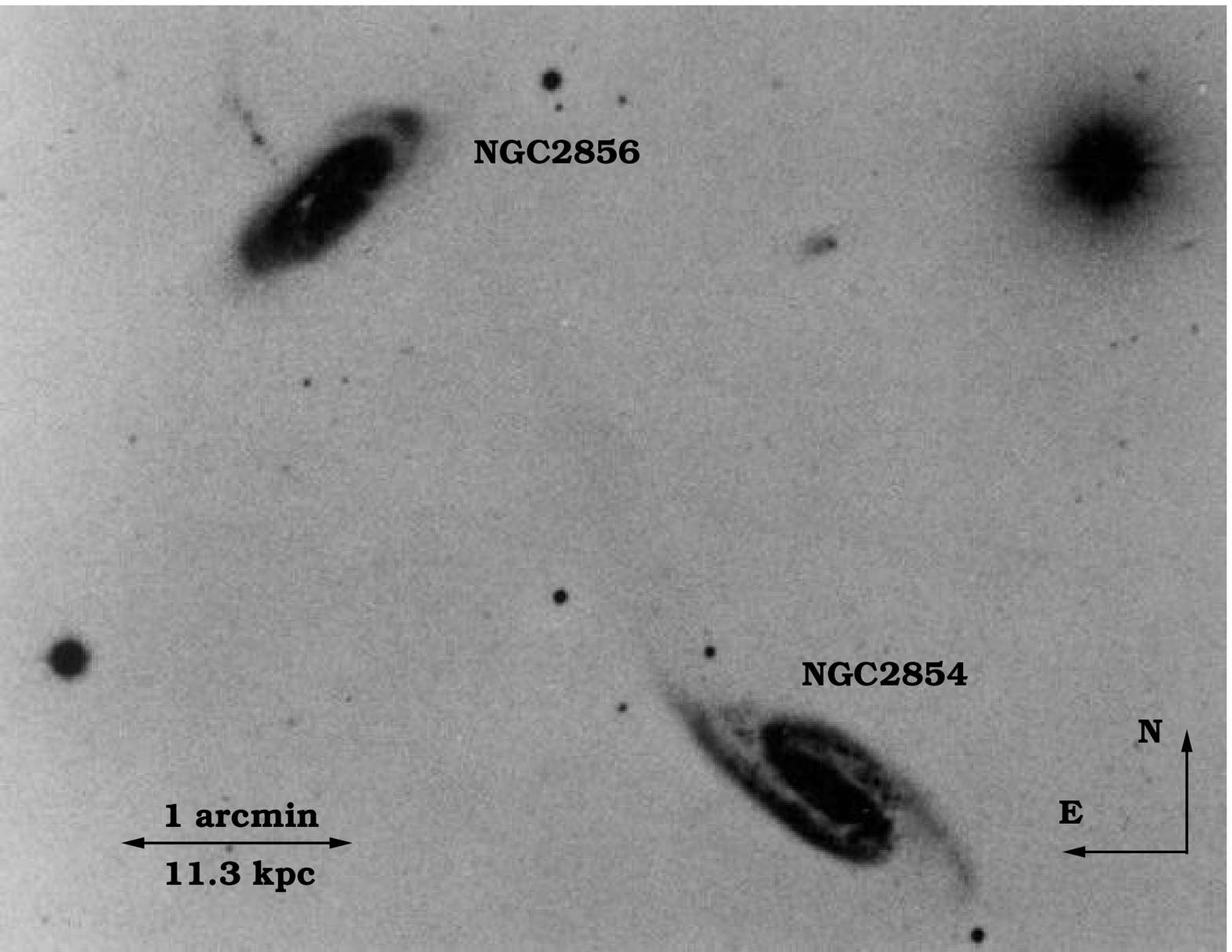}
\caption{
  \small 
The Arp (1966) image of the wide galaxy pair Arp 285 (NGC 2856/4).  
The northern galaxy NGC 2856 has an unusual
`tail' like feature extending out perpendicular to the disk
from the middle of the disk.
\citet{toomre72}
suggested that this is 
material from the bridge/companion,
which is accreting onto NGC 2856.  Note that this tail
appears clumpy in this image.
Note the `bright spot' in the northwestern edge of the NGC 2856 disk.
}
\end{figure*}

 \begin{figure*}
\caption{
  \small 
A montage of the GALEX, SDSS, and Spitzer images of NGC 2856,
the northern galaxy
in Arp 285. North is up and east to the left.
The field of view is 1\farcm1 $\times$ 0\farcm9.
Notice the series of clumps in the northern tail.
The tail clumps are enclosed by 1\farcs61 black circles in the last panel 
on the g image, and labeled as in Table 2.  
The NGC 2856 disk clumps listed in Table 2 are marked in the last panel
by 4$''$ radius white circles.  
}
\end{figure*}

\clearpage

 \begin{figure}
\caption{
  \small 
Upper left: The SARA H$\alpha$ map of NGC 2856.
Upper right: The SARA R band map of NGC 2856.
Lower left: The SARA H$\alpha$ map of
NGC 2856 (contours) superimposed on the Spitzer
8 $\mu$m map (greyscale).
Lower right: The SARA H$\alpha$ map of NGC 
2856 (contours) plotted on the SDSS g map
(greyscale).
North is up and east to the left.
This map has been smoothed by a Gaussian with FWHM = 4\farcs5.
The field of view is 1\farcm0 $\times$ 0\farcm9.
}
\end{figure}

 \begin{figure}
\caption{
  \small 
An approximately true color multi-filter optical SDSS image of NGC 2856.
Note that the clumps in the tail are blue, and the northeastern spiral
arm is bluer than the southwestern disk.
The field of view is 1\farcm1 $\times$ 0\farcm9.
}
\end{figure}

 \begin{figure}
\caption{
  \small 
The smoothed SDSS g image of NGC 2856 (color), with
21 cm HI contours (from \citealp{chengalur94}) superimposed.
North is up and east to the left.
Notice the bridge connecting
this galaxy to its southern companion NGC 2854.
Also note the bend in the northern tail north of the 
clumps marked in Figure 2.
The field of view is 3\farcm0 $\times$ 3\farcm3.
The HI beamsize is 29$''$ $\times$ 29$''$, and the HI
contours are (2.6, 4.6, 8.1, 14, 25, 43, and 76) $\times$
10$^{20}$ cm$^{-2}$.
}
\end{figure}

 \begin{figure*}
\caption{
  \small 
A montage of images of the southern galaxy in Arp 285, NGC 2854.
North is up and east to the left.
The field of view is 1\farcm2 $\times$ 1\farcm0.
Notice the series of clumps in the northern spiral arm.
In the UV and optical, the southern end of the bar is brighter than the
northern end.  At longer wavelengths, the disk is more symmetrical.
The positions of the 
8 $\mu$m-selected clumps in Table 2 are circled on the 8 $\mu$m
image in the last panel.
The circles have 2\farcs8 radii.
}
\end{figure*}

\clearpage

 \begin{figure}
\caption{
  \small 
Upper left: The SARA H$\alpha$ map of NGC 2854.
Upper right: The SARA R band map of NGC 2854.
Lower left: The SARA H$\alpha$ map of
NGC 2854 (contours) superimposed on the Spitzer
8 $\mu$m map (greyscale).
Lower right: The SARA H$\alpha$ map of NGC 
2854 (contours) plotted on the SDSS g map
(greyscale).
North is up and east to the left.
This map has been smoothed by a Gaussian with FWHM = 4\farcs5.
The field of view is 1\farcm1 $\times$ 1\farcm1.
}
\end{figure}

 \begin{figure}
\caption{
  \small 
An approximately true color multi-filter optical SDSS image of NGC 2854.
North is up and east to the left.
Note that the clumps in the northern arm are blue, and the southeastern
end of the bar is bluer than the northern end.
The southern arm/tail is also bluer than that in the north.
The field of view is 1\farcm2 $\times$ 1\farcm0.
}
\end{figure}

 \begin{figure}
\caption{
  \small 
The smoothed g image of NGC 2854, with 21 cm HI contours
from \citet{chengalur94} superimposed.
North is up and east to the left.
The field of view is 3\farcm4 $\times$ 4\farcm2.
Note the long tidal tail extending 1\farcm8 to the south.
The HI beamsize is 29$''$ $\times$ 29$''$, and the HI
contours are (2.6, 4.6, 8.1, 14, 25, 43, and 76) $\times$
10$^{20}$ cm$^{-2}$.
}
\end{figure}

 \begin{figure}
\plotone{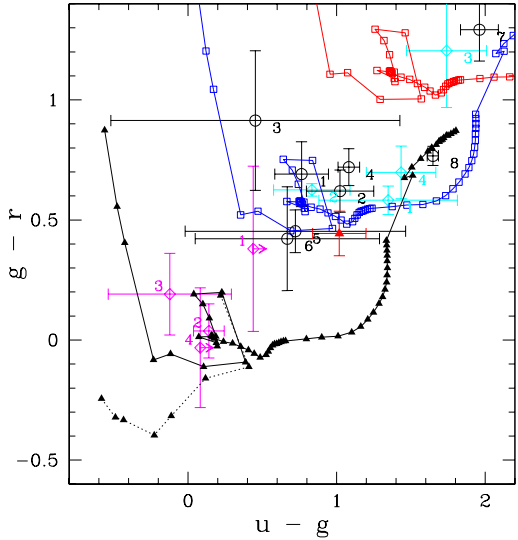}
\caption{
  \small 
The 
g $-$ r 
vs.\ 
u $-$ g 
colors of the 
clumps in the NGC 2856 tail (magenta open diamonds), the
NGC 2856 disk (cyan open diamonds), and the NGC 2854 disk (black open circles).
The clumps are identified by their numbers in Table 2, labeled in
the same color.
These data are compared
with solar metallicity Kroupa IMF instantaneous burst
population synthesis models with extinction of E(B $-$ V) = 0
(black filled triangles),
0.5 (blue open squares), and 
1.0 (red open squares).
To show the effect of H$\alpha$ on the g $-$ r color, for the zero
extinction model two curves are shown: with (solid line) and 
without (dotted line) H$\alpha$.
The model ages start with an age
of 1 Myr for the point on the left end of the curve, increasing
by 1 Myr steps to 20 Myr, then by 5 Myr steps to 50 Myrs, 10 Myr steps
to 100 Myrs, 100 Myr steps to 1 Gyr, and 500 Myr steps to 10 Gyr.
The red filled
triangle shows the colors of the 25\farcs6 $\times$ 9\farcs7 region
enclosing all four knots in the northern tail.
}
\end{figure}

 \begin{figure}
\plotone{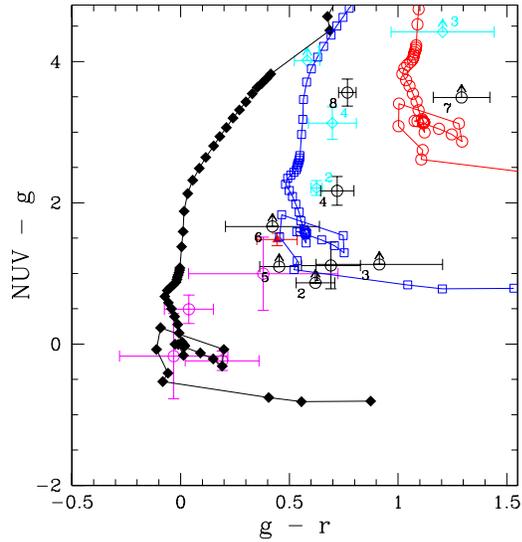}
\caption{
  \small 
The NUV $-$ g color of the 
25\farcs6 $\times$ 9\farcs7 region in the 
northern tail, plotted against g $-$ r (red filled triangle).
The clumps in the tail (magenta open diamonds),
the NGC 2856 disk (cyan open diamonds), and 
the NGC 2854 disk (black open circles) are also plotted.
Solar metallicity Kroupa IMF instantaneous burst
population synthesis model colors are also shown,
with extinction of E(B $-$ V) = 0
(black diamonds),
0.5 (blue open squares), and 1.0 (red open circles).
The model ages start with an age
of 1 Myr for the point at lower left end of the curve, increasing
by 1 Myr steps to 20 Myr, then by 5 Myr steps to 50 Myrs, then 10 Myr steps
to 100 Myrs, 100 Myr steps to 1 Gyr, and 500 Myr steps to 10 Gyr.
All models include H$\alpha$.
}
\end{figure}

 \begin{figure}
\plotone{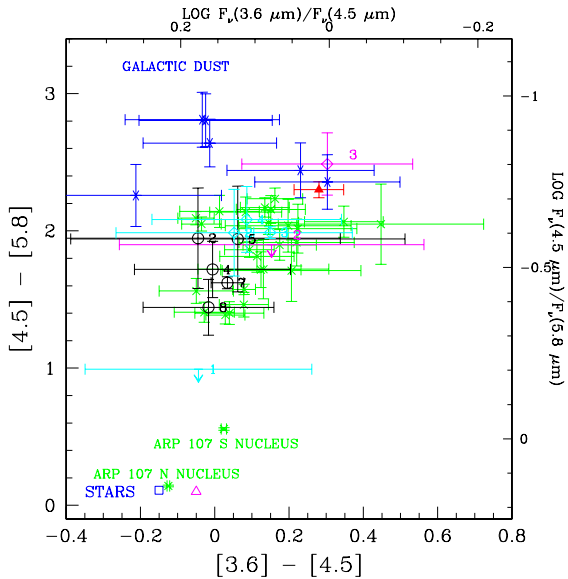}
\caption{
  \small 
The Spitzer 
[4.5] $-$ [5.8] 
vs. 
[3.6] $-$ [4.5] 
color-color
plot, showing the location of 
the clumps in the NGC 2856 tail (magenta open 
diamonds), the NGC 2856 disk (cyan open diamonds), and
the NGC 2854 disk (black open circles).
The clumps are labeled.
The colors of
M0III stars (open dark blue square), from M. Cohen (2005, private 
communication), and the mean colors of the 
field stars of \citet{whitney04} (magenta open triangle) are also shown.
The colors of normal stars all lie within 0.5 magnitudes of 0, 0 in this
plot (M. Cohen 2005, private communication).
We have also plotted 
the locations of the clumps in Arp 107
and Arp 82
as green asterisks
\citep{smith05a, hancock07},
 excluding likely
foreground stars, background quasars, upper limits,
and point with uncertainties $>$ 0.5 magnitudes.
The observed Spitzer colors \citep{flagey06} for diffuse dust towards several positions in
the Milky Way are also plotted (blue X's).
The red diamond shows the colors of the 
25\farcs6 $\times$ 9\farcs7 region
enclosing all four knots in the northern tail.
The errorbars include both statistical uncertainties and an uncertainty
in the colors due to varying the sky annuli (see text).
}
\end{figure}

 \begin{figure}
\plotone{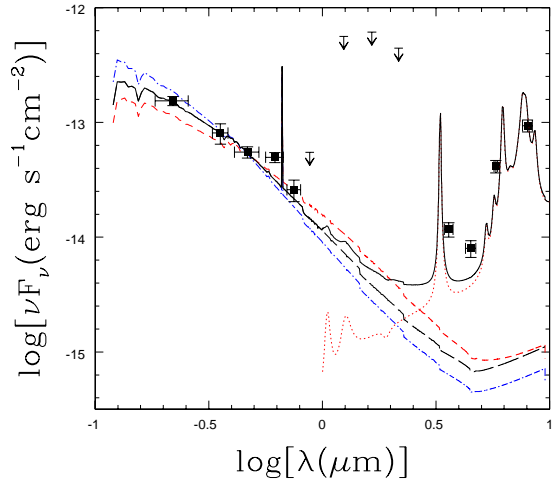}
\caption{
  \small 
The UV-mid-infrared spectral energy distribution of
clump 3 in the northern tail (filled squares), including the upper limit
in the SDSS z band (arrow).
The 2MASS upper limits are also shown.
The long dashed black curve is
the best fit solar metallicity
population synthesis model (4 Myrs, E(B$-$V)=0.1).
The dot-dashed blue and short dashed
red curves show the youngest and oldest solar metallicity
models respectively, with
their associated best-fit extinctions.
All of the models have
been normalized to the g band flux.
The red dotted curve is the \citet{dl07} Milky Way dust model
with U = 100 and q$_{PAH}$ = 4.6$\%$, scaled to the 8 $\mu$m flux.
The solid black
curve is the sum of these three components.
Note the contribution from the 3.3 $\mu$m PAH feature to the 
3.6 $\mu$m Spitzer band, and the H$\alpha$ contribution to 
the r band.   Also note that the 3.6 $\mu$m and 4.5 $\mu$m
fluxes are much higher than expected from the stellar population
synthesis model, suggesting contributions from either hot dust,
as in the dust model shown, or a second colder stellar population
undetected in the population synthesis.
}
\end{figure}

\vfill
\eject
~~~

 \begin{figure}
\plotone{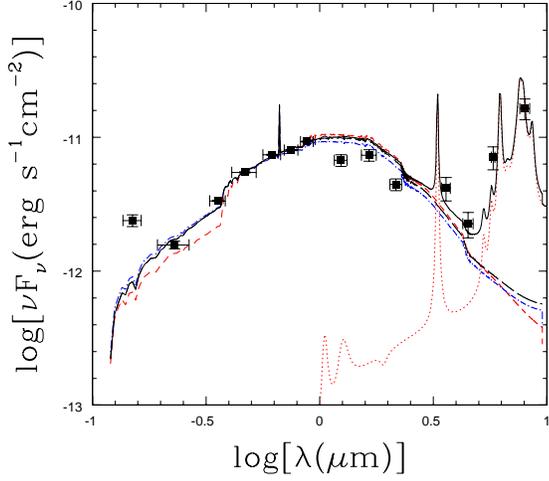}
\caption{
  \small 
The UV-mid-infrared spectral energy distribution of
clump 2 in the NGC 2856 disk.   Symbols and curves are as in Figure 13.
Note that, although the age is
similar to that of tail clump 3 (Table 4), the SED is very 
different because of the higher extinction.  Starlight contributes
a higher fraction of the 3.6 and 4.5 $\mu$m flux in this clump
than in tail clump 3.
}
\end{figure}

\newpage
\vfill
\eject

 \begin{figure}
\plotone{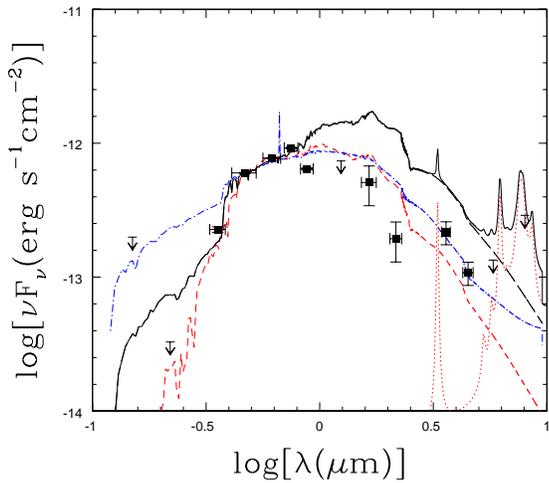}
\caption{
  \small 
The UV-mid-infrared spectral energy distribution of
the `bright spot' at the edge of the NGC 2856 disk.
Symbols and curves are as in Figure 13.
Note the upper limits at 5.8 $\mu$m and 8.0 $\mu$m, as well
as the GALEX upper limits.
Starlight can account for the 3.6 $\mu$m and 4.5 $\mu$m
emission.
}
\end{figure}

\clearpage

 \begin{figure}
\plotone{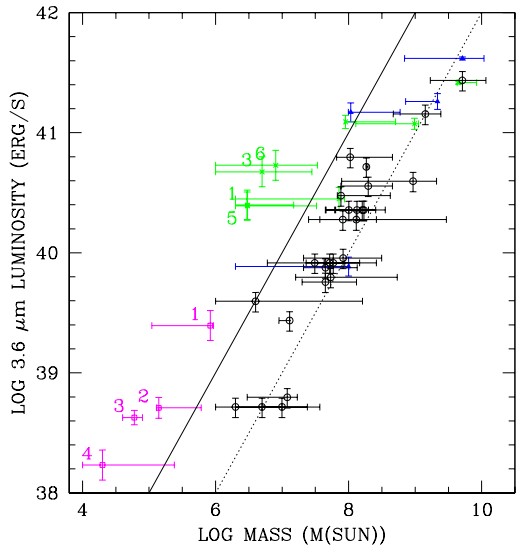}
\caption{
  \small 
The 3.6 $\mu$m luminosity of the Arp 285 clumps, compared to
their masses implied by the population synthesis models.
The magenta open squares are the tail clumps, the blue filled
triangles are 
clumps 
in the NGC 2856 disk, and the green crosses are the NGC 2854 disk
clumps.  The black open circles are the Arp 82 clumps, from
\citet{hancock07}.
A constant 
M/L$_{3.6}$ ratio of 1 M$_{\sun}$/L$_{\sun}$
is represented by the solid black line, while the dotted line
is 
M/L$_{3.6}$ = 10 M$_{\sun}$/L$_{\sun}$.
The 3.6 $\mu$m luminosity
was calculated assuming the FWHM of the bandpass 
$\Delta$$\nu$
of 
16.23 $\times$ 10$^{12}$ Hz.   
}
\end{figure}

\clearpage

\begin{figure*}
\caption{
\small 
Snapshots of the model gas disks. The stellar appearance is similar.
Red particles originated in the primary galaxy,
green in the companion.
The top left panel shows a time near closest approach (T = 0 Myrs).
The companion has swung in from the lower left, and 
swings around to an apogalacticon point at later times. 
The upper right panel and the lower left are at times 
near the present (T = 370 Myrs and 510 Myrs, respectively). 
The lower right panel is at a 
later time (T = 740 Myrs), when the companion begins to fall 
back to merge with the primary. In the first
three panels every third gas particle is plotted with a dot. 
In the second and third panels blue astericks 
mark star-forming particles, except those within 5 kpc of
the primary center, which were omitted for clarity. 
The star forming region in the northern tail was produced
from gas accreted from the companion, while the star forming
regions in the central region of the northern galaxy
were produced from gas that originated in the northern galaxy.
The star forming regions in the bridge, southern galaxy,
and southern tail were formed from gas that originated in the southern galaxy.
In the final panel only every fifth particle was plotted, to 
show the persistent spiral in the primary disk.
The motion of the companion around the point of greatest separation
is very slow, so little positional change is evident in
the last three panels.  Later timesteps are shown in the Appendix.
}
\end{figure*}

\clearpage


\clearpage


\clearpage


\clearpage








%
%
\begin{deluxetable}{ccccccccccccccc}
\tabletypesize{\scriptsize}
\def\et#1#2#3{${#1}^{+#2}_{-#3}$}
\tablewidth{0pt}
\tablecaption{Total Galaxian Magnitudes$^a$\label{tab-1}}
\tablehead{
\multicolumn{1}{c}{ID} &
\colhead{FUV} & 
\colhead{NUV} & 
\colhead{u} & 
\colhead{g} & 
\colhead{r} & 
\colhead{i} & 
\colhead{z} & 
\colhead{F$_{3.6 {\mu}m}$} & 
\colhead{F$_{4.5 {\mu}m}$} & 
\colhead{F$_{5.8 {\mu}m}$} & 
\colhead{F$_{8.0 {\mu}m}$} &
\colhead{F$_{24 {\mu}m}$} 
\\ 
\multicolumn{1}{c}{} &
\colhead{(mag)} & 
\colhead{(mag)} & 
\colhead{(mag)} & 
\colhead{(mag)} & 
\colhead{(mag)} & 
\colhead{(mag)} & 
\colhead{(mag)} 
& \multicolumn{1}{c}{(mJy)} 
& \multicolumn{1}{c}{(mJy)} 
& \multicolumn{1}{c}{(mJy)} 
& \multicolumn{1}{c}{(mJy)} 
& \multicolumn{1}{c}{(mJy)} 
\\ 
}
\startdata
NGC 2856  &    17.17         &   16.56  & 14.76  & 13.38   & 12.70    & 12.32  &     12.11 & 62.0 & 42.0 & 130.4 & 367.5  & 622.5 \\
NGC 2854 & 16.49 & 15.95 & 14.76 & 13.49 & 12.88 & 12.53 & 12.33 & 41.4 & 26.2 & 73.9 & 189.9 & 184.4 \\
ANON$^b$ & 20.91 & 20.78 & 19.96 & 18.99 & 18.99 & 18.71 & 18.70 & 0.12 & 0.08 &
$<$0.17 & $<$0.15 & $<$0.60 \\
\\
\enddata
\tablenotetext{a}{The statistical uncertainties in the optical and UV
magnitudes are typically $\sim$0.01 magnitudes.  The Spitzer
uncertainties are as given in \citet{smith07}.}
\tablenotetext{b}{At 9$^{\rm h}$ 24$^{\rm m}$ 2.9$^{\rm s}$, +49$^{\circ}$ 14$'$
41$''$ (J2000).}
\end{deluxetable}

\clearpage
%
%
\begin{deluxetable}{cccccccccccccc}
\rotate
\tabletypesize{\scriptsize}
\setlength{\tabcolsep}{0.03in}
\def\et#1#2#3{${#1}^{+#2}_{-#3}$}
\tablewidth{0pt}
\tablecaption{Magnitudes and Flux Densities for Clumps in Arp 285$^a$\label{tab-2}}
\tablehead{
\multicolumn{1}{c}{ID} &
\multicolumn{1}{c}{R.A.} &
\multicolumn{1}{c}{Dec.} &
\colhead{FUV} &
\colhead{NUV} &
\colhead{u} & 
\colhead{g} & 
\colhead{r} & 
\colhead{i} & 
\colhead{z} & 
\colhead{F$_{3.6 {\mu}m}$} & 
\colhead{F$_{4.5 {\mu}m}$} & 
\colhead{F$_{5.8 {\mu}m}$} & 
\colhead{F$_{8.0 {\mu}m}$} 
\\ 
\multicolumn{1}{c}{} &
\multicolumn{1}{c}{(J2000)} &
\multicolumn{1}{c}{(J2000)} &
\colhead{(mag)} & 
\colhead{(mag)} & 
\colhead{(mag)} & 
\colhead{(mag)} & 
\colhead{(mag)} & 
\colhead{(mag)} & 
\colhead{(mag)} 
& \multicolumn{1}{c}{(mJy)} 
& \multicolumn{1}{c}{(mJy)} 
& \multicolumn{1}{c}{(mJy)} 
& \multicolumn{1}{c}{(mJy)} 
\\ 
}
\startdata
\multicolumn{14}{c}{Clumps in the Northern NGC 2856 Tail}\\
\hline
  1  &   9 24 17.7 &   49 15  6.2 & $--$ &       22.7 $\pm$  0.3  & $\ge$22.18          &       21.74 $\pm$   0.20  &       21.36 $\pm$   0.22  &       20.98 $\pm$   0.25  & $\ge$20.45          & $\le$ 0.084          & $\le$ 0.059          & $\le$ 0.14          & $\le$ 0.47        \\
  2  &   9 24 18.2 &   49 15 12.5 & $--$ &       21.2 $\pm$  0.1  &       20.87 $\pm$   0.09  &       20.73 $\pm$   0.04  &       20.69 $\pm$   0.06  &       20.87 $\pm$   0.11  &       20.72 $\pm$   0.34  &        0.017 $\pm$   0.004  &        0.013 $\pm$   0.004  & $\le$ 0.05          & $\le$ 0.13        \\
  3  &   9 24 18.5 &   49 15 18.7 & $--$ &       21.3 $\pm$  0.1  &       21.44 $\pm$   0.22  &       21.56 $\pm$   0.12  &       21.37 $\pm$   0.11  &       21.88 $\pm$   0.23  & $\ge$20.87          &        0.014 $\pm$   0.002  &        0.012 $\pm$   0.002  &        0.08 $\pm$   0.01  &        0.25 $\pm$   0.03\\
  4  &   9 24 18.7 &   49 15 22.9 & $--$ &       22.0 $\pm$  0.1  & $\ge$22.24          &       22.16 $\pm$   0.10  &       22.19 $\pm$   0.18  &       22.27 $\pm$   0.28  & $\ge$20.89          & $\le$ 0.006          & $\le$ 0.005          & $\le$ 0.03          &        0.10 $\pm$   0.03\\
\hline
\multicolumn{14}{c}{25\farcs8 $\times$ 9\farcs7 Region Including Clumps in Northern Tail$^b$}\\
\hline
&  9 24 18.2 & 49 15 14.5 &
20.07 $\pm$  0.12  &  20.09 $\pm$  0.07  &  19.63 $\pm$  0.17  &  18.61 $\pm$  0.06  &  18.17 $\pm$  0.07  &  17.83 $\pm$  0.10  &  18.20 $\pm$  0.29  &   0.35 $\pm$  0.02  &   0.29 $\pm$  0.01  &   1.57 $\pm$  0.05  &   3.66 $\pm$  0.11 \\
\hline
\multicolumn{14}{c}{Clumps in the NGC 2856 Disk}\\
\hline
1$^c$ & 9 24 14.2 & 49 15 16.1 & $\ge$21.38   & $\ge$22.98   & 20.32 $\pm$ 0.05 & 18.97 $\pm$ 0.01 & 18.39 $\pm$ 0.01 & 17.99 $\pm$ 0.01 & 18.20 $\pm$ 0.02 & 0.260 $\pm$ 0.051 & 0.162 $\pm$ 0.032 & $\le$0.26   & $\le$0.77  \\
2 & 9 24 15.4 & 49 15 4.9 & 18.68 $\pm$ 0.11 & 18.78 $\pm$ 0.07 & 17.40 $\pm$ 0.01 & 16.57 $\pm$ 0.01 & 15.94 $\pm$ 0.01 & 15.64 $\pm$ 0.01 & 15.30 $\pm$ 0.01 & 5.008 $\pm$ 1.000 & 3.397 $\pm$ 0.730 & 13.74 $\pm$ 2.70 & 43.83 $\pm$ 7.80\\
3$^d$ & 9 24 16.1 & 49 14 57.0 & $\ge$19.86   & $\ge$21.00   & 18.32 $\pm$ 0.01 & 16.58 $\pm$ 0.01 & 15.38 $\pm$ 0.01 & 14.93 $\pm$ 0.01 & 14.50 $\pm$ 0.01 & 14.085 $\pm$ 0.421 & 10.451 $\pm$ 0.289 & 42.38 $\pm$ 1.19 & 135.01 $\pm$ 3.51\\
4 & 9 24 16.7 & 49 14 48.7 & 20.17 $\pm$ 0.18 & 19.76 $\pm$ 0.08 & 18.06 $\pm$ 0.01 & 16.63 $\pm$ 0.01 & 15.93 $\pm$ 0.01 & 15.73 $\pm$ 0.01 & 15.36 $\pm$ 0.02 & 6.178 $\pm$ 1.005 & 4.325 $\pm$ 0.733 & 19.14 $\pm$ 2.73 & 61.65 $\pm$ 7.91\\
\hline
\multicolumn{14}{c}{Clumps in the NGC 2854 Disk}\\
\hline
1 & 9 24 1.5 & 49 12 15.5 & $\ge$19.82   & 20.79 $\pm$ 0.31 & 20.44 $\pm$ 0.09 & 19.67 $\pm$ 0.09 & 18.98 $\pm$ 0.08 & 18.80 $\pm$ 0.10 & 18.87 $\pm$ 0.11 & $\le$0.847   & $\le$0.548   & 1.94 $\pm$ 0.58 & 5.79 $\pm$ 1.66\\
2 & 9 24 2.0 & 49 12 1.1 & $\ge$19.66   & $\ge$20.28   & 20.43 $\pm$ 0.07 & 19.41 $\pm$ 0.06 & 18.79 $\pm$ 0.06 & 18.49 $\pm$ 0.06 & 17.96 $\pm$ 0.06 & 0.953 $\pm$ 0.234 & 0.591 $\pm$ 0.150 & 2.31 $\pm$ 0.51 & 7.21 $\pm$ 1.42\\
3 & 9 24 2.3 & 49 12 23.1 & $\ge$20.67   & $\ge$21.45   & 20.78 $\pm$ 0.07 & 20.32 $\pm$ 0.04 & 19.41 $\pm$ 0.05 & 19.50 $\pm$ 0.06 & 19.65 $\pm$ 0.07 & $\le$1.602   & $\le$1.074   & $\le$3.28   & $\le$8.47  \\
4 & 9 24 2.4 & 49 12 11.2 & 19.37 $\pm$ 0.24 & 19.03 $\pm$ 0.15 & 17.94 $\pm$ 0.02 & 16.86 $\pm$ 0.02 & 16.14 $\pm$ 0.02 & 15.84 $\pm$ 0.03 & 15.55 $\pm$ 0.04 & 4.169 $\pm$ 0.556 & 2.683 $\pm$ 0.377 & 8.49 $\pm$ 1.08 & 25.85 $\pm$ 2.98\\
5 & 9 24 2.9 & 49 12 0.3 & $\ge$19.83   & $\ge$20.38   & 20.01 $\pm$ 0.08 & 19.29 $\pm$ 0.06 & 18.83 $\pm$ 0.06 & 18.94 $\pm$ 0.06 & 18.69 $\pm$ 0.06 & 0.833 $\pm$ 0.254 & 0.570 $\pm$ 0.160 & 2.21 $\pm$ 0.48 & 6.80 $\pm$ 1.43\\
6 & 9 24 2.9 & 49 12 25.5 & 20.54 $\pm$ 0.34 & $\ge$21.52   & 20.52 $\pm$ 0.10 & 19.85 $\pm$ 0.11 & 19.43 $\pm$ 0.13 & 19.19 $\pm$ 0.13 & 18.89 $\pm$ 0.15 & $\le$1.815   & $\le$1.218   & $\le$3.46   & $\le$9.51  \\
7$^d$ & 9 24 3.1 & 49 12 15.2 & $\ge$19.84   & $\ge$21.00   & 19.47 $\pm$ 0.06 & 17.51 $\pm$ 0.03 & 16.21 $\pm$ 0.02 & 15.53 $\pm$ 0.02 & 15.04 $\pm$ 0.01 & 8.881 $\pm$ 0.257 & 5.927 $\pm$ 0.166 & 17.15 $\pm$ 0.50 & 50.58 $\pm$ 1.40\\
8 & 9 24 3.7 & 49 12 18.3 & 20.22 $\pm$ 0.34 & 20.55 $\pm$ 0.11 & 18.63 $\pm$ 0.03 & 16.99 $\pm$ 0.02 & 16.22 $\pm$ 0.03 & 15.86 $\pm$ 0.04 & 15.59 $\pm$ 0.04 & 4.019 $\pm$ 0.451 & 2.560 $\pm$ 0.300 & 6.28 $\pm$ 0.92 & 20.53 $\pm$ 1.93\\
\enddata
\tablenotetext{a}{Except where noted, only statistical uncertainties are included, calculated from the rms in the smaller sky annulus used.}
\tablenotetext{b}{Uncertainties include both statistical uncertainties and uncertainties due to sky subtraction, calculated as in \citet{smith07}.}
\tablenotetext{c}{`Bright spot' in northwestern edge of disk.}
\tablenotetext{d}{Nucleus.}
\end{deluxetable}

%
%
\begin{deluxetable}{c|c|c|c|c}
\tabletypesize{\scriptsize}
\setlength{\tabcolsep}{0.03in}
\def\et#1#2#3{${#1}^{+#2}_{-#3}$}
\tablewidth{0pt}
\tablecaption{Parameters for Clump Photometry for Arp 285\label{tab-3}}
\tablehead{
\colhead{Telescope} &
\multicolumn{1}{c}{Aperture} &
\colhead{Aperture} & 
\multicolumn{1}{c}{Inner} &
\multicolumn{1}{c}{Outer} \\ 
\colhead{} &
\multicolumn{1}{c}{Radius} &
\colhead{Correction} & 
\multicolumn{1}{c}{Sky} &
\multicolumn{1}{c}{Sky} \\ 
\colhead{} &
\colhead{} & 
\colhead{(mag)} & 
\multicolumn{1}{c}{Annulus} &
\multicolumn{1}{c}{Annulus} \\
}
\startdata
\multicolumn{5}{c}{Clumps in the Northern NGC 2856 Tail}\\
\hline
GALEX&2 pix (3\farcs0)&0.45/0.69$^a$&2$-$5 pix (3$''$ $-$ 7\farcs5)&4 $-$ 7 pix  (6$''$ $-$ 10\farcs5)\\
SDSS&5 pix (2\farcs0)&0.08$^b$&10$-$20 pix (4$''$ $-$ 8$''$)&15 $-$ 30 pix (6$''$ $-$12$''$)\\
Spitzer&2 pix (2\farcs4)&0.21,0.23,0.35,0.50$^c$&2$-$8 pix (2\farcs4$-$9\farcs6)&5$-$9 pix (6$''$ $-$ 10\farcs8) \\
2MASS&3 pix (3\farcs0)&0.1$^b$& 4 $-$ 7 pix (4\farcs0 $-$ 7\farcs0) \\
\hline
\multicolumn{5}{c}{Clumps in the NGC 2856 Disk}\\
\hline
GALEX&2 pix (3\farcs0)&0.45/0.69$^a$&2$-$5 pix (3$''$ $-$ 7\farcs5)&4 $-$ 7 pix (6$''$ $-$ 10\farcs5)\\
SDSS&10 pix (4$''$) &0.03$^b$&10$-$ 20 pix (4$''$ $-$ 8$''$)&15 $-$ 30 pix (6$''$ $-$ 12$''$)\\
Spitzer&3 pix (3\farcs6) &0.13,0.13,0.15,0.23$^c$&5 $-$ 10 pix (6$''$ $-$ 12$''$)&9 $-$ 12 pix (10\farcs8 $-$ 14\farcs4)\\
2MASS&3 pix (3\farcs0)&0.1$^b$& 4 $-$ 7 pix (4\farcs0 $-$ 7\farcs0) \\
\hline
\multicolumn{5}{c}{Clumps in the NGC 2854 Disk}\\
\hline
GALEX&2 (3\farcs0)&0.45/0.69$^a$&2 $-$ 5 pix (3$''$ $-$ 7\farcs5)&4 $-$ 7 pix (6$''$ $-$ 10\farcs5)\\
SDSS&7 pix (2\farcs8) &0.03$^b$&10$-$ 20 pix (4$''$ $-$ 8$''$)&15 $-$ 30 pix (6$''$ $-$ 12$''$)\\
Spitzer&3 pix (3\farcs6) &0.13,0.13,0.15,0.23$^c$&5 $-$ 10 pix (6$''$ $-$ 12$''$)&9 $-$ 12 pix (10\farcs8$-$14\farcs4)\\
2MASS&3 pix (3\farcs0)&0.1$^b$&4 $-$ 7 pix (4\farcs0 $-$ 7\farcs0)\\
\hline
\enddata
\tablenotetext{a}{For FUV and NUV, respectively. NUV from bright stars in 
field. FUV from bright stars in the Arp 65 field.}
\tablenotetext{b}{From bright stars in the field.}
\tablenotetext{c}{For 3.6, 4.5, 5.8, \& 8.0 $\mu$m, respectively.
From the IRAC Data Manual, Version 3.}
\end{deluxetable}

%
%
\begin{deluxetable}{cccc}
\def\et#1#2#3{${#1}^{+#2}_{-#3}$}
\tablewidth{0pt}
\tablecaption{Model Ages and Extinctions for Clumps with Solar Metallicity Models$^a$\label{tab-4}}
\tablehead{
\multicolumn{1}{c}{Clump} &
\colhead{Age} & 
\colhead{E(B$-$V)}&
\colhead{Colors Used}\\
\colhead{} & 
\colhead{(Myr)} & 
\colhead{(mag)}&
\colhead{} \\
}
\startdata
\multicolumn{4}{c}{NGC 2856 Tail Clumps}\\
\hline
1&18 $\pm$ $^{149}_{15}$& 0.3 $\pm$ 0.3 & NUV $-$ g, g $-$ r, r $-$ i \\
2&7 $\pm$ $^{42}_{2}$ & 0.1 $\pm$ $^{0.2}_{0.1}$ & NUV $-$ g, u $-$ g, g $-$ r, r $-$ i, i $-$ z\\
3&4 $\pm$ $^{2}_{1}$ & 0.1 $\pm$ 0.1 & NUV $-$ g, u $-$ g, g $-$ r, r $-$ i\\
4& 15 $\pm$ $^{8}_{12}$ & 0.0 $\pm$ $^{0.2}_{0.0}$ & NUV $-$ g, g $-$ r, r $-$ i\\
\hline
\multicolumn{4}{c}{NGC 2856 Disk Clumps}\\
\hline
1$^b$&142 $\pm$ $^{1358}_{137}$ & 0.5 $\pm$ $^{0.2}_{0.5}$ &  u $-$ g, g $-$ r, r $-$ i\\
2& 6 $\pm$ $^{2}_{1}$ & 0.68 $\pm$ $^{0.04}_{0.1}$ & NUV $-$ g, u $-$ g, g $-$ r, r $-$ i, i $-$ z\\
3$^c$& 6900 $\pm$ $^{3200}_{6843}$ & 0.2 $\pm$ $^{0.7}_{0.1}$ & u $-$ g, g $-$ r, r $-$ i, i $-$ z\\
4 & 88 $\pm$ $^{281}_{38}$ & 0.6 $\pm$ $^{0.1}_{0.3}$ &  NUV $-$ g, u $-$ g, g $-$ r, r $-$ i, i $-$ z\\
\hline
\multicolumn{4}{c}{NGC 2854 Disk Clumps}\\
\hline
1 & 8 $\pm$ $^{4}_{5}$ & 0.4 $\pm$ $^{0.4}_{0.2}$ &  
NUV $-$ g, u $-$ g, g $-$ r, r $-$ i, i $-$ z\\
2 & 7 $\pm$ $^{2393}_{2}$ &0.6 $\pm$ $^{0.2}_{0.6}$ & 
u $-$ g, g $-$ r, r $-$ i, i $-$ z\\
3 & 4 $\pm$ $^5_3$ & 0.7 $\pm$ $^{0.4}_{0.5}$ & 
u $-$ g, g $-$ r, r $-$ i, i $-$ z\\
4&7 $\pm$ $^{1}_{2}$ & 0.6 $\pm$ $^{0.2}_{0.1}$ & 
u $-$ g, g $-$ r, r $-$ i\\
5 & 6 $\pm$ $^{1394}_{2}$ & 0.5 $\pm$ $^{0.2}_{0.3}$ &
u $-$ g, g $-$ r, r $-$ i, i $-$ z\\
6 & 36 $\pm$ $^{3164}_{32}$ & 0.4 $\pm$ $^{0.5}_{0.3}$ &
u $-$ g, g $-$ r, r $-$ i, i $-$ z\\
7$^c$ & 7 $\pm$ $^{7893}_{2}$ & 1.3 $\pm$ $^{0.1}_{1.0}$ &
u $-$ g, g $-$ r, r $-$ i\\
8 & 1200 $\pm$ $^{5500}_{259}$ & 0.2 $\pm$ $^{0.1}_{0.2}$ &
u $-$ g, g $-$ r, r $-$ i, i $-$ z \\

\enddata
\tablenotetext{a}{All clump ages obtained with instantaneous
burst models.  
The oldest
models run were 10 Gyrs old.}
\tablenotetext{b}{`Bright spot' in northwestern edge of disk.
The upper limit on the NUV flux further constrains this age to
$^{>}_{\sim}$400 Myrs (see Figure 14 and Section 5).}
\tablenotetext{c}{Nucleus}
\end{deluxetable}

%
%
\begin{deluxetable}{c|c|c|c|c}
\tabletypesize{\scriptsize}
\setlength{\tabcolsep}{0.03in}
\def\et#1#2#3{${#1}^{+#2}_{-#3}$}
\tablewidth{0pt}
\tablecaption{Optical Absolute Magnitude Ranges for Various Objects\label{tab-6}}
\tablehead{
\colhead{Class Object} &
\multicolumn{1}{c}{M$_B$} &
\colhead{M$_V$} & 
\multicolumn{1}{c}{M$_R$} &
\multicolumn{1}{c}{Notes and References} \\ 
}
\startdata
Arp 285 Tail Clumps&$-$12.0 to $-$10.6&$-$12.3 to $-$10.8&$-$12.3 to $-$10.9&This work$^a$\\
NGC 2856 `Bright Spot'&$-$13.6&$-$14.3&$-$15.0&This work$^a$\\
Other Arp 285 Disk Clumps&$-$15.9 to $-$12.1&$-$17.1 to $-$13.2&
$-$17.8 to $-$13.3&This work$^a$\\
Nearby Dwarf Irregular Galaxies&$-$18 to $-$8; median=$-$13.2&&&
\citet{kar04}\\
Local Group Irregular Leo A&$-$11.3&&&\citet{kar04}\\
Local Group Irregular GR 8&$-$12.0&&&\citet{kar04}\\
M81 Dwarf A&$-$12.4&&&\citet{patterson96}\\
Tidal Dwarf Galaxies&$-$17 to $-$12.5&&&{\it b}\\
Arp 82 Tail Clumps&&&$-$15.9 to $-$13.9&\citet{hancock07}\\
Super Star Clusters&&$-$16 to $-$12&&{\it c}\\
30 Dor R136 Star Cluster in LMC&&$-$11.3&&\citet{oconnell94}\\
\hline
\enddata
\tablenotetext{a}{Calculated using the SDSS color
transformations for stars given in \citet{jester05}. Does not include
correction for internal extinction.
}
\tablenotetext{b}{\citet{weilbacher03} and \citet{higdon06}.
Fainter end of NGC 5291 TDG range extrapolated from high end assuming
constant B $-$ [3.6] color. }
\tablenotetext{c}{\citet{holtzman92, holtzman96, oconnell94, whitmore93,
whitmore95, schweizer96, watson96}.}
\end{deluxetable}

%
%
\begin{deluxetable}{c|c|c}
\tabletypesize{\scriptsize}
\setlength{\tabcolsep}{0.03in}
\def\et#1#2#3{${#1}^{+#2}_{-#3}$}
\tablewidth{0pt}
\tablecaption{Stellar Mass Ranges for Various Objects\label{tab-7}}
\tablehead{
\colhead{} &
\colhead{Mass} &
\multicolumn{1}{c}{Notes and References} \\ 
}
\startdata
Arp 285 Tail Clumps&2 $\times$ 10$^{4}$ $-$ 8 $\times$ 10$^5$ M$_{\sun}$&
This work$^a$\\
NGC 2856 `Bright Spot'&$\sim$10$^8$ M$_{\sun}$&This work$^a$\\
Other Non-Nuclear Arp 285 Disk Clumps&3 $\times$ 10$^6$ $-$ 10$^9$ M$_{\sun}$
&This work$^a$\\
Arp 285 Nuclei&5 $\times$ 10$^9$ M$_{\sun}$&This work$^a$\\
Dwarf Irregular Galaxies&6.3 $\times$ 10$^6$ $-$ 1.2 $\times$ 10$^9$ M$_{\sun}$; Median = 10$^8$ M$_{\sun}$&\citet{hunter85}$^b$\\
Tidal Dwarf Galaxies&2 $\times$ 10$^6$ $-$ 5 $\times$ 10$^8$ M$_{\sun}$&{\it c}\\
Arp 82 Tail Clumps&5 $\times$ 10$^6$ $-$ 8 $\times$ 10$^7$ M$_{\sun}$ &{\it d}\\
Local Group Globular Clusters&3 $\times$ 10$^3$ $-$ 3 $\times$ 10$^6$ M$_{\sun}$; Median = 1.3 $\times$ 10$^5$ M$_{\sun}$&
\citet{mclaughlin05}\\
\hline
\enddata
\tablenotetext{a}{The uncertainties on these masses are a factor of 2 $-$ 5.
}
\tablenotetext{b}{Converted to H$_{\rm 0}$ = 75 km~s$^{-1}$Mpc$^{-1}$.
}
\tablenotetext{c}{\citet{braine01} and \citet{higdon06}.
}
\tablenotetext{d}{\citet{hancock07}. 
Uncertainties are a factor of $\sim$2 $-$ 10.}\\
\end{deluxetable}



\begin{thebibliography}{}

\bibitem[Abazajian et al.(2003)]{abazajian03}
Abazajian, K., et al.\ 2003, AJ, 126, 2081

\bibitem[Abraham \& van den Bergh(2001)]{abraham01}
Abraham, R. G., \& van den Bergh, S. 2001, Science, 293, 1273

\bibitem[Arp(1966)]{arp66}
Arp, H. 1966, Atlas of Peculiar Galaxies (Pasadena: Caltech)

\bibitem[Barnes \& Hernquist(1992)]{barnes92}
Barnes, J. E., \& Hernquist, L. 1992, Nature, 360, 715

\bibitem[Bastian et al.(2005)]{bastian05}
Bastian, N., Gieles, M., Efremov, Y. N.,
\&
Lamers, H. J. G. L. M.
2005, A\&A, 443, 79


\bibitem[Bloemen et al.(1986)]{bloemen86}
Bloemen, J. B. G. M., et al. 1986, A\&A, 154, 25

\bibitem[Bohlin, Savage, \& Drake(1978)]{bohlin78}
Bohlin, R. C., Savage, B. D., \& Drake, J. F.
1978, ApJ, 224, 132

\bibitem[Braine et al.(2001)]{braine01}
Braine, J., Duc, P.-A., Lisenfeld, U., Charmandaris, V., 
Vallejo, O., Leon, S., \& Brinks, E. 2001, 
A\&A, 378, 51

\bibitem[Bushouse, Lamb, \& Werner(1988)]{bushouse88}
Bushouse, H. A., Lamb, S. A., \& Werner, M. W. 1988, 
ApJ, 335, 74

\bibitem[Calzetti et al.(2005)]{calzetti05}
Calzetti, D., et al.\ 2005, ApJ, 633, 871

\bibitem[Calzetti, Kinney, \& Storchi-Bergmann(1994)]{calzetti94}
Calzetti, D., Kinney, A. L., \& Storchi-Bergmann, T. 1994, ApJ, 429, 582

\bibitem[Cervi\~no et al.(2002)]{cervino02}
Cervi\~no, M., Valls-Gabaud, D., Luridiana, V., \& Mas-Hess, J. M.
2002, A\&A, 381, 51

\bibitem[Chen \& Wu(2007)]{chen07}
Chen, C. \& Wu, H. 2007, AJ, 133, 1710

\bibitem[Chengalur, Salpeter, \& Terzian(1994)]{chengalur94}
Chengalur, J. N., Salpeter, E. E., \& Terzian, Y. 1994, AJ, 107, 1984

\bibitem[Chengalur, Salpeter, \& Terzian(1995)]{chengalur95}
Chengalur, J. N., Salpeter, E. E., \& Terzian, Y. 1995, AJ, 110, 167

\bibitem[Cox(2000)]{cox00}
Cox, A. N., Allen's Astrophysical Quantities, 4th Edition (Springer).


\bibitem[Cutri et al.(2006)]{cutri06}
Cutri, R. M., et al.\ 2006, Explanatory Supplement to the 
2MASS All Sky Release and Extended Mission Products,
http://www.ipac.caltech.edu/2mass/

                                                                                              
\bibitem[Dale et al.(2005)]{dale05}
Dale, D. A., et al.\ 2005, ApJ, 633, 857
                                                                                              
\bibitem[Draine \& Li(2007)]{dl07}
Draine, B. T., \& Li, A. 2007, ApJ, 657, 810

\bibitem[Duc \& Mirabel(1994)]{duc94}
Duc, P.-A. \& Mirabel, I. F. 1994, A\&A, 289, 83

\bibitem[Duc et al.(1997)]{duc97}
Duc, P.-A., Brinks, E., Wink, J. E., \& Mirabel, I. F. 1997, A\&A, 326, 537

\bibitem[Elmegreen \& Efremov(1996)]
{elmegreen96}
Elmegreen, B. G. \& Efremov, Y. N.
1996, ApJ, 466, 802

\bibitem[Elmegreen, Kaufman, \& Thomasson(1993)]{elmegreen93}
Elmegreen, B. G., Kaufman, M., \& Thomasson, M.
1993, ApJ, 412, 90

\bibitem[Elmegreen et al.(2006)]{elmegreen06}
Elmegreen, D. M., Elmegreen, B. G., Kaufman, M., Sheth, K., Struck, C.,
Thomasson, M., \& Brinks, E. 2006, ApJ, 642, 158

\bibitem[Fazio et al.(2004)]{fazio04}
Fazio, G. G., et al.\ 2004, ApJS, 154, 10

\bibitem[Feldmeier et al.(2002)]{feldmeier02}
Feldmeier, J. J. et al.\ 2002, ApJ, 575, 779

\bibitem[Flagey et al.(2006)]{flagey06}
Flagey, N., Boulanger, F., Vrestraete, L.,
Miville Desch\^enes, M. A.,
Noriega Crespo, A., \& Reach, W. T.
2006, A\&A, 453, 969

\bibitem[Hancock et al.(2007)]{hancock07}
Hancock, M., Smith, B. J., Struck, C., Giroux, M. L., Appleton, P. N.,
Charmandaris, V., \& Reach, W. T. 2007, AJ, 133, 676



\bibitem[Hibbard \& van Gorkom(1996)]{hibbard96}
Hibbard, J. E. \& van Gorkom, J. H. 1996, AJ,
111, 655

\bibitem[Higdon, Higdon, \& Marshall(2006)]{higdon06}
Higdon, S. J. U., Higdon, J. L., \& Marshall, J. 2006, ApJ, 640
768

\bibitem[Holtzman et al.(1992)]{holtzman92}
Holtzman, J. A., et al.\ 1992, AJ, 103, 691
                                                                                
\bibitem[Holtzman et al.(1996)]{holtzman96}
Holtzman, J. A., et al.\ 1996, AJ, 112, 416

\bibitem[Hunter \& Gallagher(1985)]{hunter85}
Hunter, D. A., \& Gallagher, J. S. III, 1985, ApJS, 58, 533

\bibitem[Jester et al.(2005)]{jester05}
Jester, S., et al.\ 2005, AJ, 130, 873

\bibitem[Karachentsev et al.(2004)]{kar04}
Karachentsev, I. D., Karachentseva, V. E., 
Huchtmeier, W. K., \&
Makarov, D. I. 2004,
AJ, 127, 2031

\bibitem[Kennicutt et al.(1987)]{kennicutt87}
Kennicutt, R. C., Jr., et al.\ 1987, AJ, 93, 1011.

\bibitem[Kroupa(2002)]{kroupa02}
Kroupa, P. 2002, Science, 295, 85

\bibitem[Larsen(2004)]{larsen04}
Larsen, S. S. 2004, A\&A, 416, 537

\bibitem[Leitherer et al.(1999)]{leitherer99}
Leitherer, C., et al. 1999, ApJS, 123, 3

\bibitem[Li et al.(2007)]{li07}
Li, H., Wu, H., Cao, C., \& Zhu, Y. 2007, AJ, 134, 1315

\bibitem[Longmore et al.(1979)]{longmore79}
Longmore, A. J., Hawarden, T. G., Cannon, R. D., Allen, D. A.,
Mebold, U., \& Goss, W. M. 1979, MNRAS, 188, 285

\bibitem[MacArthur et al.(2004)]{macarthur04}
MacArthur, L. A., Courteau, S., Bell, E., \& Holtzman, J. A. 2004, 
ApJS, 152, 175

\bibitem[Martin et al.(2005)]{martin05}
Martin, D. C., et al.\ 2005, ApJ, 619, L1

\bibitem[McLaughlin \& van der Marel(2005)]{mclaughlin05}
McLaughlin, D., \& van der Marel, R. P. 2005, ApJ, 161, 304

\bibitem[Morris \& van den Bergh(1994)]{morris94} 
Morris, S. L., \& van den Bergh, S. 1994, ApJ, 427, 696

\bibitem[O'Connell, Gallagher, \& Hunter(1994)]{oconnell94}
O'Connell, R. W., Gallagher, J. S., \& Hunter, D. A. 1994, ApJ, 433, 65

\bibitem[Pasquali, de Grijs, \& Gallagher(2003)]{pas03}Pasquali, A. de Grijs, R., \& Gallagher, J. S.,
2003, \mnras, 345, 161

\bibitem[Patterson \& Thuan(1996)]{patterson96}
Patterson, R. J., \& Thuan, T. X. 1996, ApJS, 107, 103

\bibitem[Press et al.(1992)]{press92}
Press, W. H., Teukolsky, S. A., Vetterling, W. T., \& Flannery, B. P.
1992, Numerical Recipes in Fortran, Second Edition (Cambridge
University Press, Cambridge), p. 692.


\bibitem[Reshetnikov et al.(2006)]{resh06}
Reshetnikov, V., 
Bournaud, F., Combe, F., 
Fa\'undez-Abans, M., \& de Oliveira-Abans, M.
2006, A\&A, 446, 447

\bibitem[Rieke et al.(2004)]{rieke04}
Rieke, G. H., et al.\ 2004, ApJS, 154, 25

\bibitem[Roche(2007)]{roche07}
Roche, N. 2008, RMxAA, in press (astro-ph/0605015)

\bibitem[Sanders et al.(1988)]{sanders88}
Sanders, D. B., Soifer, B. T., Elias, J. H.,
Madore, B. F., Matthews, K., Neugebauer, G., 
\& Scoville, N. Z. 1988, ApJ, 325, 74

\bibitem[Schweizer et al.(1996)]{schweizer96}
Schweizer, F., Miller, B. W., Whitmore, B. C., \& Fall, S. M. 1996, AJ, 112, 1839


\bibitem[Smith et al.(1987)]{smith87}
Smith, B. J., Kleinmann, S. G., Huchra, J. P., \&
Low, F. 1987, ApJ, 318, 161

\bibitem[Smith, Struck, \& Pogge(1997)]{smith97}
Smith, B. J., Struck, C., \& Pogge, R. W. 1997, ApJ, 483, 754

\bibitem[Smith et al.(2005a)]{smith05a}
Smith, B. J., Struck, C., \& Nowak, M. A. 2005a, AJ,
129, 1350


\bibitem[Smith et al.(2005b)]{smith05b}
Smith, B. J., Struck, C., Appleton, P. N., Charmandaris, V., 
Reach, W., \& Eitter, J. J. 2005b, AJ, 130, 2117

\bibitem[Smith et al.(2007)]{smith07}
Smith, B. J., Struck, C., Hancock, M., Appleton, P. N., Charmandaris, V., 
\&
Reach, W. 2007, AJ, 133, 791

\bibitem[Soifer et al.(1987)]{soifer87}
Soifer, B. T., 
Sanders, D. B., Madore, B. F., Neugebauer,
G., Danielson, G. E., Elias, J. H.,
Lonsdale, C. J., \& Rice, W. L. 1987,
ApJ, 320, 238

\bibitem[Struck(1997)]{struck97}
Struck, C. 1997, ApJS, 113, 269

\bibitem[Struck \& Smith(2003)]{struck03}
Struck, C. \& Smith, B. J. 2003, ApJ, 589, 157

\bibitem[Struck et al.(2005)]{struck05}
Struck, C., 
Kaufman, M., Brinks, E., Thomasson, M., Elmegreen, B. G.,
\& Elmegreen, D. M.
2005, MNRAS, 364, 69

\bibitem[Toomre \& Toomre(1972)]{toomre72}
Toomre, A. \& Toomre, J. 1972, ApJ, 178, 623


                                                                                                       
\bibitem[V\'{a}zquez \& Leitherer(2005)]{vaz05}V\'{a}zquez, G. A. \& Leitherer, C. 2005, ApJ, 621, 695

\bibitem[Wallin \& Struck-Marcell(1988)]{wallin88}
Wallin, J. F. \& Struck-Marcell, C. 1988, AJ,
96, 1850


\bibitem[Watson et al.(1996)]{watson96}
Watson, A. M., et al.\ 1996, AJ, 112, 534

\bibitem[Weilbacher, Duc, \& Fritze-v.\ Alvensleben(2003)]{weilbacher03}
Weilbacher, P. M., Duc, P. A., \& Fritze-v.\ Alvensleben, U. 2003,
A\&A, 397, 545

\bibitem[Werner et al.(2004)]{werner04}
Werner, M. W., et al.\ 2004, ApJS, 154, 1

\bibitem[Whitmore \& Schweizer(1995)]{whitmore95}
Whitmore, B. C. \& Schweizer, F. 1995, AJ, 109, 960

\bibitem[Whitmore et al.(1993)]{whitmore93}
Whitmore, B. C., Schweizer, F., Leitherer, C., Borne, K., \& Robert, C.
1993, AJ, 106, 1354

\bibitem[Whitney et al.(2004)]{whitney04}
Whitney, B. A., et al.\ 2004, ApJS, 154, 315

\bibitem[York et al.(2000)]{york00}
York, D. G., et al.\ 2000, AJ, 120, 1579

\bibitem[Zhang, Fall, \& Whitmore(2001)]{zhang01}
Zhang, Q., Fall, S. M., \& Whitmore, B. C. 2001,
ApJ, 561, 727


\end{thebibliography}
\end{document}